\documentclass[times,twocolumn]{aastex631}

\usepackage[flushleft]{threeparttable}
\usepackage{amsmath}
\usepackage{multirow}
\usepackage{bm}
\usepackage{pbox}
\usepackage{xcolor}

\shortauthors{Lin et al.}

\begin{document}

\title{Early-Universe-Physics Insensitive and Uncalibrated Cosmic Standards: \\ Constraints on $\Omega_{\rm{m}}$ and Implications for the Hubble Tension}

\author[0000-0003-2240-7031]{Weikang Lin}
\email{weikanglin@sjtu.edu.cn}
\affiliation{Tsung-Dao Lee Institute, Shanghai Jiao Tong University, Shanghai 200240, China}
\affiliation{Physics Department, North Carolina State University, Raleigh, NC 27695, USA}

\author[0000-0002-1071-6526]{Xingang Chen}
\email{xingang.chen@cfa.harvard.edu}
\affiliation{Institute for Theory and Computation, Harvard-Smithsonian Center for Astrophysics, 60
Garden Street, Cambridge, MA 02138, USA}

\author[0000-0001-8927-1795]{Katherine J. Mack}
\email{kmack@ncsu.edu}
\affiliation{Physics Department, North Carolina State University, Raleigh, NC 27695, USA}



\begin{abstract}

To further shed light on whether pre-recombination models can resolve the Hubble tension, we explore constraints on the cosmic background evolution that are insensitive to early-universe physics. The analysis of the cosmic microwave background (CMB) anisotropy has been thought to highly rely on early-universe physics. However, we show that the fact that the sound horizon at recombination being close to that at the end of the drag epoch is insensitive to early-universe physics. This allows us to link the absolute sizes of the two horizons and treat them as free parameters. Jointly, the CMB peak angular size, Baryon Acoustic Oscillations (BAO), and Type Ia supernovae can be used as ``early-universe-physics insensitive and uncalibrated cosmic standards'', which measure the cosmic history from recombination to today. They can set strong and robust constraints on the post-recombination cosmic background, especially the matter density parameter with $\Omega_{\rm{m}}=0.302\pm0.008$ ($68\%$ C.L.) assuming a flat $\Lambda$CDM after recombination. When we combine these with other non-local observations, we obtain several constraints on $H_0$ with significantly reduced sensitivity to early-universe physics. These are all more consistent with the Planck 2018 result than the local measurement results such as those based on Cepheids. This suggests a tension between the post-recombination, but non-local, observations and the local measurements which cannot be resolved by modifying pre-recombination early universe physics.

\end{abstract}



\section{Introduction}
The standard cosmological model---a spatially flat universe with a cosmological constant and cold dark matter ($\Lambda$CDM)---has been successful overall in explaining and predicting cosmological observations \citep{2018-Planck-cosmo-params,2020-SDSS-IV,2019-DES-year1,Scolnic-etal-2018-Pantheon,Cooke-etal-2018}. However, recently several tensions on cosmological parameters measured via different observations have been reported to various extents; see, e.g.,  \citet{2021-Riess-etal-H0,2020-KiDS-1000,KiDS-VIKING-DES-2019}. The currently most hotly debated tension is the Hubble tension: the difference in the Hubble parameter between the results from methods based on $\Lambda$CDM (such as the Planck CMB anisotropy and/or BAO) and the results from the method of the distance ladder (especially the Cepheid-based local measurement) \citep{2021-Riess-etal-H0,2018-Planck-cosmo-params,2021-DiValentino-etal-H0-review}.

It is important to explore different methods to measure or infer $H_0$ \citep{2017-Moresco-Marulli-CC,Dominguez-etal-2019,2017-GW-170717-H0,H0LiCOW-2019}, because this helps us conclude whether the Hubble tension has an origin beyond the standard model or which systematic effects need more investigation. Although most methods of determining $H_0$ depend on an underlying cosmological model, the reliance varies and the constraints exhibit different degeneracy directions, e.g., in the $H_0$-$\Omega_{\rm{m}}$ plane. Based on the standard $\Lambda$CDM model, it was shown that most constraints from different types of observations consistently overlap on some common parameter region in the $H_0$-$\Omega_{\rm{m}}$ plane \citep{Lin_Mack_Hou_2020}, which to some extent disfavors a beyond-the-standard-model explanation.  

Nonetheless, the Hubble tension is often described as a discrepancy between the early and the late\footnote{In this work, ``early'' refers to cosmic time before recombination and ``late'' refers to cosmic time later than that.} universe \citep{Verde-etal-2019}. This is partially because the two powerful methods with CMB and BAO assume the standard pre-recombination physics and partially because observations such as Type Ia supernovae (SNe Ia) and BAO (even without assuming the standard pre-recombination physics) strongly constrain late-time deviations from $\Lambda$CDM \citep{Aylor-etal-2019,Evslin-etal-2018-price-to-shift-H0,2020-Krishnan-Colgain-Sheikh}. It is suggested that models that shorten the early expansion history and the sound horizon scale are promising (or ``the least unlikely'') \citep{2020-Knox-Millea}. A number of examples of this class of models have been proposed, such as a form of dark energy that appeared between matter-radiation equality and recombination \citep{Poulin-etal-2019-EDE,Agrawal-2019lmo,2020-Jeremy-Mark-early-dark-neutrino,2020-Niedermann-Sloth,2020-Smith-Poulin-Amin-oscillating-scalar-field}, dark matter that partially interacts with dark radiation \citep{2016-Chacho-Cui-Yanou,2020-Choi-Yanagida-Yokozaki}, neutrinos with self interaction \citep{Kreisch-etal-2019-self-interacting-neutrinos,2020-Das-Ghosh} or interaction with dark matter \citep{2020-Ghosh-Khatri-Roy}, early modified gravity \citep{2020-Braglia-Ballardini-Finelli-Koyama}, and the consideration of primordial magnetic field \citep{Pogosian-Jedamzik2020}. Works are being undertaken to confront those proposals with different observations, and currently it still remains controversial whether or not they are viable solutions to the Hubble tension \citep{Hill-etal-2020,2020-Smith-etal,Philcox-etal-2020,2020-Ivanov-McDonough-Hill-early-DE-lss,2020-Jedamzik-Pogosian-Zhao,2020-Haridasu-Balakrishna-Viel,2020-Choudhury-Hannestad-Tram,DAmico:2020ods,2020-Achidiacono-Gariazzon-Giunti-Hannestad,2020-Brinckmann-Hyeok-Loverde,2017-Raveri-Hu-Hoffman-Wang-partialy-acoustic-DM}.

Given the effort devoted to the early-universe resolutions to the Hubble tension, it is especially important to explore methods that determine $H_0$ independent of or insensitive to the early-universe physics. There are already several methods that only involve post-recombination physics such as the cosmic chronometers \citep{Jimenez-2002-chronometers}, measuring the redshift dependence of $\gamma$-ray optical depth \citep{2013-Dominguez-Prada-gamma-ray-H0}, the strong-lensing time-delay technique \citep{1964-Refsdal}, inferring cosmic age from old stars \citep{2019-Jimenez-etal-stellar-and-cosmic-age}, and the gravitational wave multi-messenger method \citep{2017-GW-170717-H0}. But currently the precision on $H_0$ from those methods is relatively low because of their own astrophysical uncertainties and their background degeneracy in the $H_0$-$\Omega_{\rm{m}}$ plane. To mitigate the latter factor, some of those methods (especially the cosmic chronometers and the $\gamma$-ray optical depth) are combined with other observations to break the background degeneracy \citep{2017-Moresco-Marulli-CC,Dominguez-etal-2019}. However, in doing so, either the constraining power of the other observation is not strong enough (e.g., combining with the uncalibrated SNe Ia), or a dependence on early-universe physics (e.g., combining with BAO or a CMB prior on $\Omega_{\rm{m}}$ from Planck) is introduced. An important aim in this work is to develop a method that provides a strong constraint on the cosmic background evolution to break the degeneracy with other observations and yet is insensitive to early-universe physics (i.e., the dependence on early-universe physics is significantly reduced). The dependence on early-universe physics may be traded with that on other potentially unaccounted-for astrophysics effect. Thus, we investigate a number of independent probes that are different in methodology. 

It is well known that, when the sound horizon scale is treated as a free parameter, it is degenerate with $H_0$, but the Hubble-distance normalized cosmic background evolution can still be constrained \citep{Aylor-etal-2019}. Several recent works have applied ``uncalibrated''\footnote{Being ``uncalibrated'' means that the absolute magnitude or scale of the standard candles or rulers are not determined by calibration with another astrophysical observation or calculated with an assumed model.} BAO and SNe Ia jointly with large-scale structure to obtain sound-horizon independent constraints on $H_0$ \citep{2020-Baxter-Sherwin,2020-Philcox-Sherwin-Farren-Baxter,2020-Pogosian-Zhao-Jedamzik}. But uncalibrated BAO does not provide a strong enough constraint on, e.g., $\Omega_{\rm{m}}$ compared to that from Planck. 

On the other hand, information obtained from the CMB has been thought to depend strongly on early-universe physics. However, as we will show, the sound horizon at the recombination epoch ($r_*$), related to the angular sizes of the CMB acoustic peaks \citep{2018-Planck-cosmo-params}, has a property that significantly reduces the model sensitivity: namely, its {\em difference} (normalized by Hubble distance) from the sound horizon at the end of the baryonic-drag epoch\footnote{The end of the drag epoch refers to the time when photon pressure is no longer able to prevent baryons from falling into the potential wells of the cold dark matter.} ($r_{\rm{d}}$) is small, which we denote as $\Delta r H_0$.
Both horizons share a common expression before recombination, making their difference to nearly cancel out the dependence on early-universe physics. Furthermore, while the redshifts of these two epochs differ by only $\Delta z_{\rm{s}}\approx 30$, the redshifts at which these two epochs imprinted observable effects---peaks in CMB power spectra and BAO extracted from two-point correlation functions of tracers of the underlying matter density fluctuation at large scales---differ by $\Delta z_{\rm{s}} \sim 1000$. The former narrow redshift separation makes $\Delta r H_0$ both very small and insensitive even to non-standard physics happening between the two epochs, while the latter large separation provides a long lever arm to effectively constrain model parameters (such as $\Omega_{\rm{m}}$) once the two angular scales are used jointly.

We will treat the sound horizon as a free parameter for CMB and jointly analyzes the CMB acoustic peak angular scale, $\theta_{\rm{cmb}}$, and BAO. We will show that the aforementioned tight relation between the two horizon scales allows us to jointly analyze $\theta_{\rm{cmb}}$ and BAO in a robust way insensitive to  early-universe physics. Such an analysis will be shown to have constraining power on $\Omega_{\rm{m}}$ comparable to that from Planck in a full $\Lambda$CDM-based analysis. We call such a joint analysis using $\theta_{\rm{cmb}}$, BAO and SNe Ia ``Early-Universe-Physics Insensitive and Uncalibrated Cosmic Standards" (EUPIUCS, or UCS for simplicity). 

Reanalyses of observations relaxing some strong assumptions are very important. After relaxing some assumptions, if the result is inconsistent with the original one, it would be an indication that those assumptions could be problematic. On the contrary, if the result is consistent with the original one, it would cause difficulties for nonstandard models/considerations trying to modify those assumptions to resolve a tension. 

This work is organized as follows. In Sec.\,\ref{sec:methods} we describe the methods of our joint analysis of UCS, in particular the connection between $\theta_{\rm{cmb}}$ and BAO. In Sec.\,\ref{sec:results} we present the constraint on the post-recombination background evolution, especially the strong constraint on $\Omega_{\rm{m}}$, and discuss the robustness of this analysis against any possible early-universe nonstandard physics. Then in Sec.\,\ref{sec:joint-analysis-with-other} we combine the UCS with other early-universe-physics independent or insensitive observations to break the background degeneracy and obtain several constraints of $H_0$. In Sec.\,\ref{sec:discussion}, we discuss how to generalize the analysis of UCS to test post-recombination cosmological models independent of early-universe physics. Finally, we summarize and conclude in Sec.\,\ref{se:summary-and-conclusion}. The Appendix contains some details of the analyses.

Throughout this work, we adopt units in which the speed of light is $c=1$. So, e.g., $H_0$ has dimensions of inverse distance unless its unit is explicitly specified.

\section{Methods}\label{sec:methods}
It is well known that based on the standard $\Lambda$CDM model, uncalibrated SNe Ia and (late-time) BAO can constrain $\Omega_{\rm{m}}$ \citep{Pantheon-2018,Aylor-etal-2019}. For SNe Ia, it is the relative change of the apparent brightness at different redshifts that constrains the matter fraction in the standard $\Lambda$CDM model, which determines the relative change of the Hubble parameter. For uncalibrated BAO, it is the change (at different redshifts) of the angular size or the redshift span of the sound horizon scale at the end of the drag epoch that sets the constraint. Those constraints are independent of early-universe physics. While detailed descriptions of those two analyses can be found in the literature \citep{Aylor-etal-2019,Pantheon-2018}, we provide discussions in Appendix \ref{sec:data-and-likelihoods} using our own notation to unify these two analyses as well as that of the CMB acoustic peak we shall discuss.\footnote{We also provided a fast algorithm to compute the SN Ia likelihood in Appendix \ref{sec:appendix-sn}.} Since the absolute magnitude of SNe Ia ($M_0$) and the absolute (comoving) scale of the sound horizon at the drag epoch $r_{\rm{d}}$ are both degenerate with the Hubble constant, we treat the combinations $\mathcal{M}\equiv M_0 -5\log_{10}(10\,{\rm{pc}}\times H_0)$ and $r_{\rm{d}}H_0$ as free parameters. 

The angular size ($\theta_{\rm{cmb}}$) of the CMB acoustic peaks has been measured precisely \citep{2018-Planck-cosmo-params}. Earlier, it has been proposed to use $\theta_{\rm{cmb}}$ (along with the ``drift parameter'' $R$ that we will not use) to, e.g., test late-time dark energy models \citep{1997-Bond-Efstathiou-Tegmark-1997}. But this has already assumed that the scale of the sound horizon ($r_*$) at recombination is ``calibrated by early-universe physics'' in the sense that $r_*$ is calculated assuming a standard cosmological model at early times. Like $r_{\rm{d}}$ for BAO, $r_*$ for $\theta_{\rm{cmb}}$ is also degenerate with $H_0$. Therefore, we also treat the combination $r_*H_0$ as a free parameter. But unlike BAO that has measurements at multiple redshifts, $\theta_{\rm{cmb}}$ is only measured at one single redshift, $z_*$. Therefore, when analyzed alone, $\theta_{\rm{cmb}}$ is not able to simultaneously constrain $r_*H_0$ and $\Omega_{\rm{m}}$.

The underlying principle of our analysis is the fact that $r_{\rm{d}}$ and $r_*$ are tightly related to each other. As shown in Appendix \ref{sec:link-between-bao-cmb}, by taking the difference between $r_{\rm{d}}H_0$ and $r_*H_0$ we have,
\begin{equation}\label{eq:difference-rH0}
    \Delta rH_0\equiv(r_{\rm{d}}-r_*)H_0=\int^{z_*}_{z_d}\frac{c_{\rm{s}}(z)}{E(z)}dz\,,
\end{equation}
where $z_*$ and $z_{\rm{d}}$ are the redshifts at recombination and at the end of the drag epoch, respectively, and $E(z)$ represents the normalized total energy evolution as a function of redshift (also see Appendix \ref{sec:notation-and-background} for the definition of $E(z)$). We can thus treat $r_*H_0$ as a free parameter and link $r_{\rm{d}}H_0$ to $r_*H_0$ by $\Delta rH_0$. The value of $\Delta rH_0$ is much smaller than $r_*H_0$ and this fact is almost independent of early-universe physics. Therefore, a substantial change to $\Delta rH_0$ is required in order to alter our analysis. But the value of $\Delta rH_0$ itself is quite insensitive to early-universe physics. This is because by taking the difference between $r_{\rm{d}}H_0$ and $r_*H_0$ the effects from early-universe physics get largely canceled. We discuss the robustness of these properties in Sec.\,\ref{sec:robustness-of-the-result}. 

We will first perform our default analysis where we use some standard assumptions (described below) to model and calculate $\Delta rH_0$. Later in Sec.\,\ref{sec:robustness-of-the-result}, we will quantify the sensitivity to changes of those assumptions and show that the results obtained are robust against any reasonable modification of our standard assumptions. We will also perform analyses separately using the Planck \citep{2018-Planck-cosmo-params} and the ACT+WMAP \citep{2020-Aiola-etal-ACT} results of $\theta_{\rm{cmb}}$ and the priors on $z_*$ and $\Delta z_{\rm{s}}$, in order to show that our results are robust against possible systematic errors in each CMB experiment. We assume a flat-$\Lambda$CDM universe after recombination.

\subsection{Our standard analysis}\label{sec:standard-analysis}
The standard assumptions in our default analysis are as follows. First, to calculate the sound speed, we need the reduced baryon fraction $\Omega_{\rm{b}}h^2$, for which we use the BBN constraint $\Omega_{\rm{b}}h^2=0.0222\pm0.0005$ \citep{Cooke-etal-2018}. We also need to know $z_*$ and $z_{\rm{d}}$, which we separately obtain from Planck \citep{2018-Planck-cosmo-params} ($z_*=1089.99\pm0.28$ and $z_*-z_{\rm{d}}=30.27\pm0.57$) and from ACT+WMAP \citep{2020-Aiola-etal-ACT} ($z_*=1089.91\pm0.42$ and $z_*-z_{\rm{d}}=29.95\pm0.75$). Next, as we are assuming the standard $\Lambda$CDM after recombination, the function $E(z)$ and the sound speed $c_{\rm{s}}$ take their standard forms; see \autoref{eq:Eofz} and \autoref{eq:sound-speed}. In particular, for the function $E(z)$ we consider a universe with a cosmological constant filled with pressureless matter (cold dark matter and baryons), photons, and neutrinos (one massless and two massive with a normal mass hierarchy).

\begin{figure}[tbp]
    \centering
    \includegraphics[width=\linewidth]{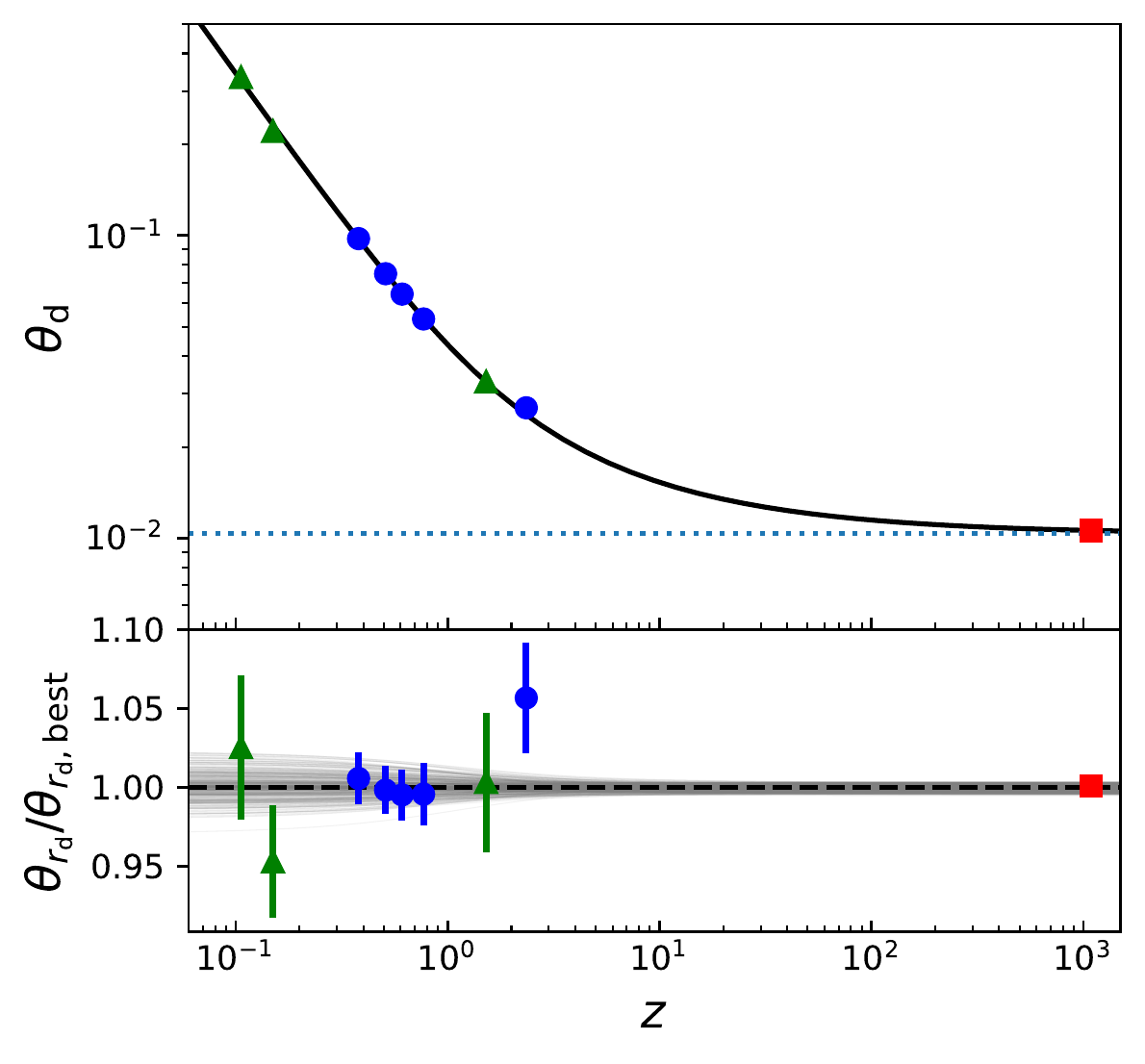}
    \caption{\label{fig:showing_thetad}The observed/inferred angular sizes $\theta_{\rm{d}}$ of the drag-epoch sound horizon at different redshifts. Top: the solid line is the prediction of $\theta_{\rm{d}}$ from the best fit model. The dotted line represents the predicted asymptotic $\theta_{\rm{d}}$ at infinite redshift. The blue circles are the observed $\theta_{\rm{d}}$'s from some late-time BAO. The green triangles are the inferred $\theta_{\rm{d}}$'s for late-time BAO that only provides $\frac{r_{\rm{d}}H_0}{f_{\rm{V}}(z)}$. The red square is the inferred $\theta_{\rm{d}}(z_*)$ from CMB. To get $\theta_{\rm{d}}(z_*)$ for the CMB, we added to $\theta_{\rm{cmb}}$ the inferred angular size difference ($\Delta\theta$) between the sound horizon at the end of the drag epoch and at recombination according to the best-fit model. The uncertainty of the inferred $\theta_{\rm{d}}(z_*)$ includes the derived uncertainty of $\Delta\theta$. The error bars are too small to show. Bottom: the ratios of the observed and predicted $\theta_{\rm{d}}$'s. The thin lines represent the predictions of the model with a subset of parameter points sampled in the MCMC analysis.}
\end{figure}

\begin{figure}[htbp]
    \includegraphics[width=\linewidth]{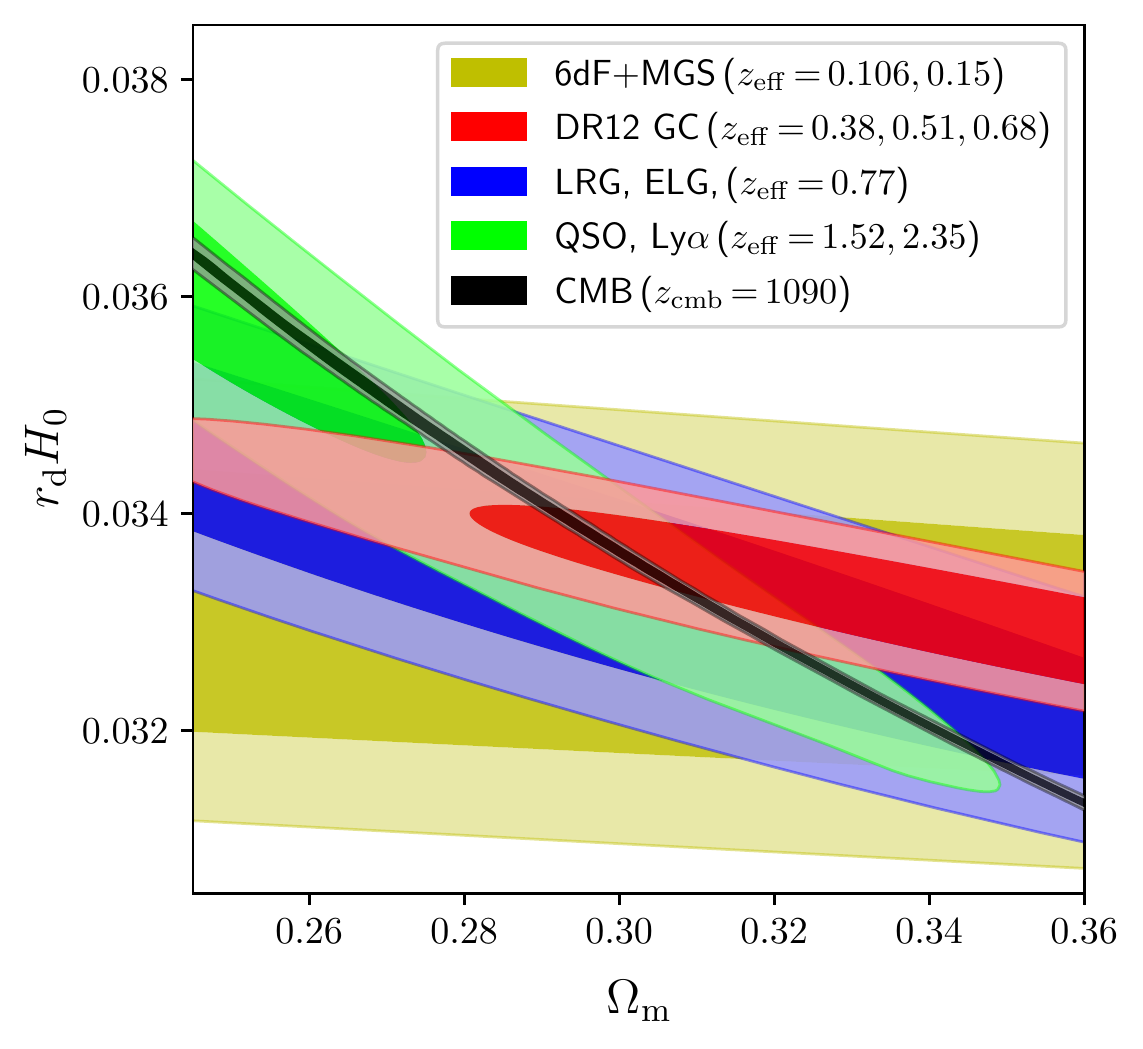}
    \caption{\label{fig:rdH0-Om}Degeneracy direction of constraints on the $\Omega_{\rm{m}}$-$r_{\rm{d}}H_0$ plane from BAO at different redshifts and $\theta_{\rm{cmb}}$. The $\theta_{\rm{cmb}}$ and BAO constraints break the degeneracy to provide a constraint on, e.g., $\Omega_{\rm{m}}$. See Table \ref{tab:BAO-measurement} for a detailed descriptions of BAOs.}
\end{figure}

The joint analysis of the $\theta_{\rm{cmb}}$ likelihood and the BAO likelihood will break the degeneracy in the $\Omega_{\rm{m}}$-$r_{\rm{d}}H_0$ plane and significantly improve the constraint on $\Omega_{\rm{m}}$ compared to the uncalibrated BAO alone. We illustrate how this works as follows. Recall that the angular size of the sound horizon at the drag epoch is given by 
\begin{equation}
    \theta_{\rm{d}}(z)=\frac{r_{\rm{d}}H_0}{f_{\rm{M}}(z)}\,,
\end{equation}
where $f_{\rm{M}}$ defined in \autoref{eq:comoving-dist-DM} is the Hubble-distance normalized comoving distance. BAO uses the angular size and the redshift span of the (drag-epoch) sound horizon, $\theta_{\rm{d}}$ and $\Delta z_{r_{\rm{d}}}$, at different effective redshifts to constrain $\Omega_{\rm{m}}$, which determines the relative change of the Hubble parameter with redshift; see Appendix \ref{sec:appendix-bao} for a discussion. With our standard assumptions, the difference $\Delta rH_0$ is $\sim6\times10^{-4}$ according to \autoref{eq:difference-rH0}, given that BAO alone can already constrain $\Omega_{\rm{m}}$ with a certain precision. With $\Delta rH_0$ roughly known, the measurement of $\theta_{\rm{cmb}}$ can be viewed as another measurement of $\theta_{\rm{d}}$ at $z_*$ (with an offset of $\Delta\theta\equiv\theta_{\rm{d}}(z_*)-\theta_{\rm{cmb}}=\frac{\Delta rH_0}{f_{\rm{M}}(z_*)}$). We plot in Figure \ref{fig:showing_thetad} the measured or inferred values of $\theta_{\rm{d}}$ at different redshifts along with the theoretical prediction according to the best-fit model\footnote{To illustrate how uncalibrated $\theta_{\rm{cmb}}$ and BAO constrain the cosmic background evolution, we only plot $\theta_{\rm{d}}$ for BAO but not $\Delta z_{r_{\rm{d}}}$. The bottom panel of our Figure \ref{fig:showing_thetad} is similar to Figure 2 in \citet{2020-SDSS-IV} except that we are plotting $\theta_{\rm{d}}$ in the $y$ axis. But we note that we are adding $\theta_{\rm{cmb}}$ in the analysis in an early-universe-physics insensitive way.}. We can see that the prediction matches the measured/inferred $\theta_{\rm{d}}$ at different redshifts quite well. The precise measurement of $\theta_{\rm{cmb}}$ (and then the inferred $\theta_{\rm{d}}$) adds a powerfully constraining anchor at a large redshift, so that the constraint on the relative change of the Hubble parameter can be significantly tightened, and hence so can the $\Omega_{\rm{m}}$ in the standard $\Lambda$CDM model. For the same (comoving) $r_{\rm{d}}$, the angular size $\theta_{\rm{d}}$ asymptotically approaches the minimum value at large redshift, as indicated in the upper panel of Figure \ref{fig:showing_thetad}. This gives the advantage that $z_*$ has to be significantly incorrect in order to affect the analysis.

BAO measurements at different redshifts give different degeneracy directions in the $\Omega_{\rm{m}}$-$r_{\rm{d}}H_0$ plane \citep{2019-Cuceu-Farr-Lemos-FontRibera}. This also applies to $\theta_{\rm{cmb}}$. In Figure \ref{fig:rdH0-Om}, we plot different constraints in the $\Omega_{\rm{m}}$-$r_{\rm{d}}H_0$ plane at different redshifts including the one inferred from $\theta_{\rm{cmb}}$. At lower redshifts, BAO measurements are more sensitive to $r_{\rm{d}}H_0$, because $f_{\rm{M}}(z)$ (and also $E(z)$) is not sensitive to $\Omega_{\rm{m}}$ at low redshifts. This is shown in Figure \ref{fig:rdH0-Om} as the constraint from 6df+MGS ($z_{\rm{eff}}<0.2$) in the $\Omega_{\rm{m}}$-$r_{\rm{d}}H_0$ plane is almost horizontal. At larger redshifts, BAO measurements are sensitive to both $r_{\rm{d}}H_0$ and $\Omega_{\rm{m}}$; this is shown in Figure \ref{fig:rdH0-Om} as constraints from BAO at higher redshifts ($z_{\rm{eff}}>0.3$). The CMB constraint is diagonal in the $\Omega_{\rm{m}}$-$r_{\rm{d}}H_0$ plane. Therefore, the tight constraint from $\theta_{\rm{cmb}}$ breaks the degeneracy in the $\Omega_{\rm{m}}$-$r_{\rm{d}}H_0$ plane of the BAO measurements at lower redshifts, and significantly improves the constraint on $\Omega_{\rm{m}}$.

To constrain $\Omega_{\rm{m}}$, besides $\theta_{\rm{cmb}}$ and uncalibrated BAO, we will also consider the Pantheon catalog of Type Ia supernovae (without calibration). We denote the joint likelihood of all these UCS data as $\mathcal{L}_{\rm{UCS}}$. In our standard analysis, the free parameters are $\Omega_{\rm{m}}$,  $r_*H_0$, $\mathcal{M}$, and $h$ (a very weak dependence; see Sec.\,\ref{sec:notation-and-background}), with some assumed prior on $z_*$, $\Delta z_{\rm{s}}$ and $\Omega_{\rm{b}}h^2$.

\section{Results}\label{sec:results}
\subsection{A tight constraint on matter density parameter insensitive to early-universe physics}
\begin{figure*}[tbp]
    \centering
    \includegraphics[width=\textwidth]{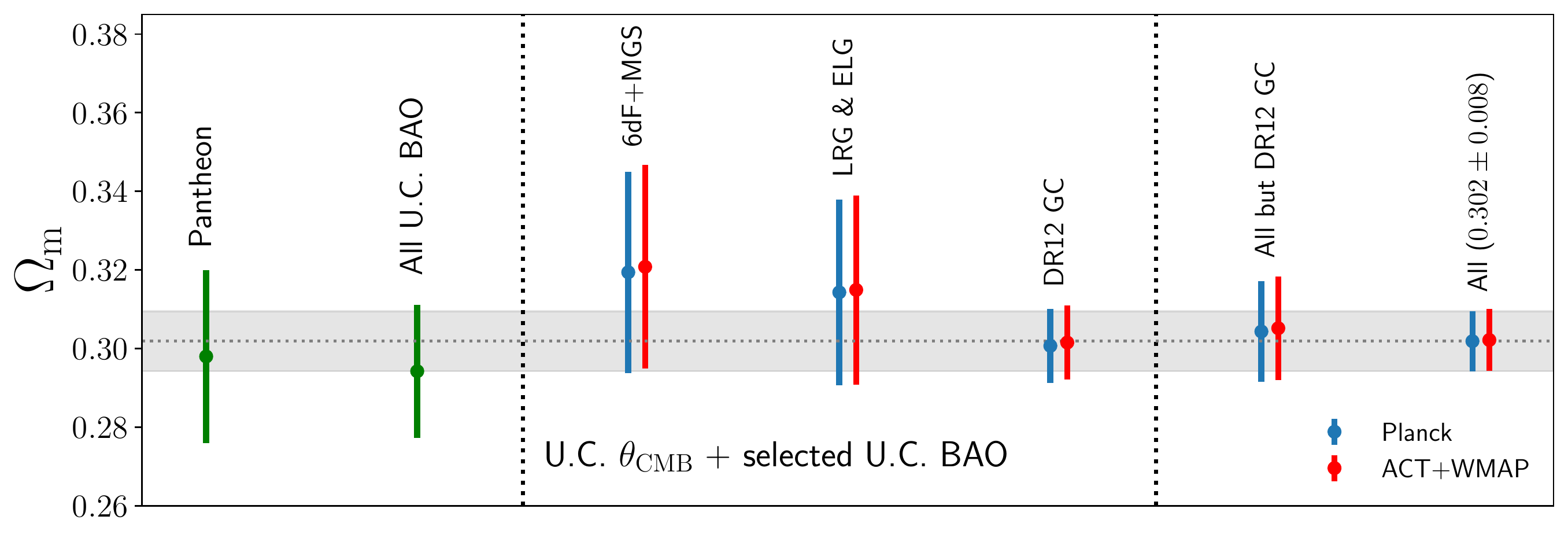}
    \caption{Constraints on $\Omega_{\rm{m}}$ obtained from different combinations of uncalibrated SN, BAO and $\theta_{\rm{cmb}}$. The two green error bars are the constraints on $\Omega_{\rm{m}}$ from Pantheon SNe Ia and all late-time BAO uncalibrated. The middle three are the results from uncalibrated $\theta_{\rm{cmb}}$ jointly with some individual late-time BAO. Different colors represent results using $\theta_{\rm{cmb}}$ from Planck and ACT+WMAP, respectively. The error bar on the right is the final result with all UCS and the second one from the right excludes the DR12 GC BAOs ($z_{\rm{eff}}=0.38,\,0.51,\,0.68$). The shaded horizontal band represents the final result ($\Omega_{\rm{m}}=0.302\pm0.008$) with Planck $\theta_{\rm{cmb}}$. All results are consistent with each other.}
    \label{fig:constraints-Om}
\end{figure*}

As mentioned earlier, even uncalibrated, standard rulers and candles can put a constraint on $\Omega_{\rm{m}}$ \citep{Pantheon-2018,Aylor-etal-2019}. With $\theta_{\rm{cmb}}$ added according to the method described above, the constraint is now much stronger. In Figure \ref{fig:constraints-Om}, we show the constraints on $\Omega_{\rm{m}}$ set by different combinations of those UCS. We first reproduced the results from Pantheon SNe Ia and uncalibrated BAO, respectively, shown by the green error bars (1-$\sigma$) in the left panel of Figure \ref{fig:constraints-Om}. In the middle panel, we show the constraints from the combination of $\theta_{\rm{cmb}}$ and three selected late-time BAO. We adopt two different $\theta_{\rm{cmb}}$ results, one from Planck temperature and polarization (TTTEEE+lowE) with \footnote{There is some correlation between $\theta_{\rm{cmb}}$ and $z_{*}$, which has also been considered in our analyses.}
\begin{equation}\label{eq:theta-from-Planck}
100\theta_{\rm{cmb}}^{\rm{Planck}}=1.04109\pm0.00030  \,,  
\end{equation}
shown in blue \citep{2018-Planck-cosmo-params}; and the other from ACT+WMAP with
\begin{equation}\label{eq:theta-from-act}
100\theta_{\rm{cmb}}^{\rm{ACT+WMAP}}=1.04170\pm0.00068\,,
\end{equation}
shown in red \citep{2020-Aiola-etal-ACT}. Among these three constraints, the one from the combination of $\theta_{\rm{cmb}}$ and the BAO from the SDSS DR12 Galaxy Clustering (DR12 GC) \citep{Alam-etal-2017} is the tightest, which is due to both the degeneracy breaking and the strong constraining power of this late-time BAO. Nonetheless, the constraints on $\Omega_{\rm{m}}$ via those different combinations are consistent with each other and with the SN-Ia-only and uncalibrated-BAO-only constraints. 

We then produce the joint constraint with all UCS. The final result is shown in the right panel of Figure \ref{fig:constraints-Om} with 
\begin{equation}\label{eq:final-constraint-on-Om}
    \Omega_{\rm{m}}=0.302\pm0.008\,,
\end{equation} using the Planck constraint on $\theta_{\rm{cmb}}$. This is a strong constraint on $\Omega_{\rm{m}}$. In fact, with the same data the full analysis in $\Lambda$CDM (assuming the standard modeling of the sound horizon) gives a constraint on $\Omega_{\rm{m}}$ that is only a little tighter, $\Omega_{\rm{\rm{m}}}=0.310\pm0.006$. Although the average is somewhat lower, our constraint on $\Omega_{\rm{m}}$ is consistent with the full analysis that assumes the standard $\Lambda$CDM model for the entire cosmic history.

Compared to the full standard analysis of CMB anisotropy, our method using UCS is only based on the geometrical information from CMB anisotropy and relies on many fewer assumptions. Firstly, as explained, our method is insensitive to early-universe physics. Secondly, our method is less vulnerable to systematic bias in the CMB observations. Indeed, in the full analysis of CMB, the matter fraction is constrained by the scale dependence of the CMB power spectra \citep{2018-Planck-cosmo-params}. This renders the constraint of $\Omega_{\rm{m}}$ correlated with, e.g., the spectral index $n_{\rm{s}}$. Incidentally, there have been discussions on the possibility of some small inconsistency between the large and small scale spectra in the Planck CMB power spectra, which might be due to unaccounted-for systematic errors or more radically some beyond-the-standard-model physics \citep{Addison-etal-2016}; but also see \citet{Planck-XI-2016} for a discussion. Our method is clearly free of those potential biases.

As we mentioned earlier, the combination of $\theta_{\rm{cmb}}$ and the DR12 GC BAO provides the strongest constraint on $\Omega_{\rm{m}}$. If there is some systematic error in the DR12 GC BAO, the result would be biased. Therefore, we also produce a joint constraint on $\Omega_{\rm{m}}$ using all UCS except for the DR12 GC, which is shown in the second to the right set of error bars in Figure \ref{fig:constraints-Om}. As we can see, even without the DR12 GC BAO, the result is fully consistent with the one using all UCS.

\subsection{Robustness against early-universe physics}\label{sec:robustness-of-the-result}
In the last subsection, we treat $r_*$ as a free parameter, which have already relaxed the most of the strong assumptions of early-universe physics. In our standard analysis, we have only made an assumption that the difference between $r_*H_0$ and $r_{\rm{d}}H_0$ can be calculated according to the standard $\Lambda$CDM model. We have assumed some priors on $z_*$ and $\Delta z_{\rm{s}}$ from CMB observations and $\Omega_{\rm{b}}h^2$ from BBN, so that we can calculate $\Delta rH_0$. One may worry that these may render our results dependent upon early-universe physics. We now show that results obtained based on this standard analysis are robust against changes caused by possible nonstandard early-universe physics. We quantify how significantly these assumptions need to be modified to substantially affect our results. We also discuss several possibilities that may be contemplated in future model-building to circumvent the arguments and constraints derived in this paper.

Recall that the underlying reason we can jointly analyze the $\theta_{\rm{cmb}}$ and uncalibrated BAO likelihoods is that the two horizon scales ($r_*$ and $r_{\rm{d}}$) are very close to each other but the measurements are so separated in redshift that together they can provide a stringent constraint on the post-recombination background evolution. We discuss this in detail in the following.

First, the fact that the difference between $r_*H_0$ and $r_{\rm{d}}H_0$ is very small is important and quite robust. Because it is small, the difference between the two angles $\theta_d(z_*)$ and $\theta_{\rm{cmb}}$ is comparable to or even smaller than the error on $\theta_d(z_*)$ extrapolated from the BAO measurements alone. Therefore, the CMB measurement on $\theta_{\rm{cmb}}$ can be viewed as another statistically significant measurement of the BAO sound horizon at a very high redshift with an uncertainty mainly caused by the uncertainty of $\Delta\theta$. 

Second, because the value of $\Omega_{\rm{m}}$ continuously and monotonically impacts the background evolution from $z\sim 1000$ to $z\sim 1$, with the same relative errors, the constraining power on $\Omega_{\rm{m}}$ from the data-derived $\theta_{\rm d}$ values grows if the redshifts of these data are further separated. 
Consequently, since $z_*$ is so high, the effect on the constraint on $\Omega_{\rm{m}}$ due to a change in $\Delta rH_0$ from nonstandard physics is further reduced. The above two points, $\Delta rH_0$ being small and $z_*$ being high, argue that our analysis is robust against the change in $\Delta rH_0$ due to possibly unknown nonstandard physics. 

Indeed, even if we artificially multiply \autoref{eq:difference-rH0} by a factor of $110\%$ mimicking some possible unknown modifications to our standard assumptions, the joint constraint on $\Omega_{\rm{m}}$ only changes to $0.303\pm0.008$, compared to $0.302\pm0.008$ in our standard analysis, or
\begin{equation}
    \frac{\delta(\Omega_{\rm{m}}^{\rm{mean}})}{\Omega_{\rm{m}}^{\rm{mean}}}=0.036\frac{\delta (\Delta rH_0)}{\Delta rH_0}\,.
\end{equation}
The small coefficient above verifies the conclusion that the constraint on $\Omega_{\rm{m}}$ is insensitive to changes to $\Delta rH_0$.

Third, even the small value of $\Delta rH_0$ itself is quite robust and it is not easy for nonstandard models to change it substantially. While the form of \autoref{eq:difference-rH0} is generally true, one may change $\Delta rH_0$ by changing the sound speed $c_{\rm{s}}$, the function $E(z)$, and the redshift span between $z_{\rm{d}}$ and $z_*$. For example, changing the $\Omega_{\rm{b}}h^2$ can change the sound speed. Regarding this, due to the consistency of the $\Omega_{\rm{b}}h^2$ constraint between Planck and BBN \citep{2018-Planck-cosmo-params,Cooke-etal-2018,2020-Fields-Olive-Yeh-Young}, and between different uses of information from CMB anisotropy \citep{2020a-Motloch}, it is difficult for the reduced baryon fraction to be significantly different from $\Omega_{\rm{b}}h^2\sim0.0222$. Since the $\Omega_{\rm{b}}h^2$ term only has a relatively small effect on $c_{\rm{s}}$ between $z_{\rm{d}}$ and $z_*$, it is even more difficult to change $\Delta rH_0$ via the $\Omega_{\rm{b}}h^2$ term. Quantitatively, even if we artificially change $\Omega_{\rm{b}}h^2$ by $10\%$ ($\sim4\sigma$ away from the mean value of our adopted prior), which we use to mimic effects from possible changes in the sound speed, $\Delta rH_0$ only changes by $\lesssim2\%$. For the $E(z)$ function, if nonstandard physics occurred before recombination as in a number of currently proposed early-universe nonstandard models, its functional form will not change.\footnote{For a late-time non-standard model, one needs to modify the standard $E(z)$ using the corresponding non-standard physics, to capture the effect on $\Delta rH_0$.} Due to the narrow redshift span between $z_{\rm{d}}$ and $z_*$, it is difficult to substantially change $\Delta rH_0$ from its standard value by changing $E(z)$ with modified energy compositions. Also, although a change to the redshift span itself can directly affect $\Delta rH_0$, to get a $\sim10\%$ change to $\Delta rH_0$, we would need a $\sim10\%$ change of $\Delta z_{\rm{s}}$, which is $\sim5.5\sigma$ away from the mean value of $\Delta z_{\rm{s}}$ inferred from Planck. It is difficult to achieve this because the decoupling between photons and baryons are based on well known physics. The redshift span between recombination and the drag epoch cannot be substantially different from the standard value unless there are some other significant interactions between, e.g., baryon and dark matter, which would have led to observable features in CMB anisotropy \citep{Dvorkin-Blum-2014,Xu-Dvorkin-Chael-2018,2018-Boddy-Bluscevic,2019-dePutter-Dore-Jerome-Green-Meyers}. As a support, \citet{2020-Jedamzik-Pogosian-Zhao} finds that $r_*\sim1.018r_{\rm{d}}$ persists in a number of pre-recombination nonstandard models. In summary, it is not easy to make a noticeable change (like $10\%$) to $\Delta rH_0$.

Besides the change of $\Delta rH_0$, one may worry about the prior on $z_*$ that we adopted. But as discussed in Appendix \ref{sec:appendix-CMB-theta}, our result is not sensitive to the prior on $z_*$, because $z_*$ is sufficiently large and $f_{\rm{M}}$ is approaching a constant (though dependent of $\Omega_{\rm{m}}$). To numerically verify this, we artificially shift the mean value of Planck result of $z_*$ from $1090$ to $1110$, which can be achieved, e.g., in the presence of primordial magnetic fields \citep{Pogosian-Jedamzik2020}. This only leads a $\sim0.2\%$ change in the mean value of $\Omega_{\rm{m}}$ (from $0.3021$ to $0.3015$).

Another possible concern is that we have adopted the constraint on $\theta_{\rm{cmb}}$ based on the standard $\Lambda$CDM model. The uncertainty of $\theta_{\rm{cmb}}$ from Planck is about $0.03$\%. Such a small uncertainty of $\theta_{\rm{cmb}}$ may not represent the possible uncertainty due to some nonstandard early-universe physics. For example, releasing the effective relativistic particle number, the Planck result becomes $100\theta_{\rm{cmb}}=1.04136\pm0.00060$. And releasing the mass of neutrinos gives $100\theta_{\rm{cmb}}=1.04105\pm0.00032$. So, the uncertainty of $\theta_{\rm{cmb}}$ due to these potential nonstandard physics at early times is comparable to or larger than $0.03$\%. However, these uncertainties in $\theta_{\rm{cmb}}$ will not substantially change our result in $\Omega_{\rm{m}}$ for the following reason. Take the combination of $\theta_{\rm{cmb}}$ and the DR12 GC BAO, for example. As shown in Figure \ref{fig:rdH0-Om}, the uncertainty of $\Omega_{\rm{m}}$ is mainly due to the width of the projection of the overlap between these two constraints in the $\Omega_{\rm{m}}$ direction. As we can visualize in that figure, even if we increase the width of the black contour or shift it a bit horizontally, the projection in the $\Omega_{\rm{m}}$ direction would not be substantially changed. Quantitatively, this insensitivity can be shown by alternatively using the result of $\theta_{\rm{cmb}}$ from ACT+WMAP. We can see from \autoref{eq:theta-from-Planck} and \autoref{eq:theta-from-act} that there is a $\sim 2\sigma$ (in terms of Planck's uncertainty) difference in the mean value of $\theta_{\rm{cmb}}$ between Planck and ACT+WMAP, and also the uncertainty of $\theta_{\rm{cmb}}$ for ACT+WMAP is more than twice as that for Planck. But Figure \ref{fig:constraints-Om} shows the resultant constraints on $\Omega_{\rm{m}}$ are nearly identical.

To evade the constraint on the post-recombination cosmic background evolution by UCS, one would need 1) some mechanism that makes the expressions for the two sound-horizon scales different prior to the recombination; or 2) some nonstandard physics occurring during the narrow gap between $z_*$ and $z_{\rm{d}}$ that significantly changes the function $E(z)$ or the sound speed $c_{\rm{s}}$; or 3) a substantial change of the redshift span between recombination and the drag epoch. None of those possibilities are easy to achieve, but nonetheless could serve as guidelines if we would like to build models of counterexamples.

Our UCS constraint on $\Omega_{\rm{m}}$ is strong and insensitive to early-universe physics. This can be used as a strong prior or be combined with other observations to break degeneracy in the background. We will do that in Sec.\,\ref{sec:joint-analysis-with-other} to obtain several early-universe-physics insensitive constraints on $H_0$.

\subsection{Constraint on the $H_0$-$\Omega_{\rm{m}}$ plane}
\label{sec:constraint-on-H0-Om}
In our standard analysis the constraint in the $H_0$-$\Omega_{\rm{m}}$ plane is actually \emph{not} exactly along the $\Omega_{\rm{m}}$ direction. That is because the normalized comoving distance $f_{\rm{M}}$ has a weak dependence on the Hubble constant via the radiation and the massive neutrino terms; see Appendix \ref{sec:notation-and-background} for a discussion. A principle component analysis shows that 
\begin{equation}\label{eq:constraint-Omh008}
    \frac{\Omega_{\rm{m}}}{0.3}\left(\frac{h}{0.7}\right)^{-0.08}=1.0060\pm0.0258\,.
\end{equation}
Therefore, the constraint exhibits a small positive correlation in the $H_0$-$\Omega_{\rm{m}}$ plane. In situations where joint analyses using MCMC with other datasets are unavailable (such as that with $\gamma$-ray optical depth that shall be discussed), we will use \autoref{eq:constraint-Omh008} as a prior to set weights in parameter chains obtained from other analyses in order to approximate the joint result. In practice, we find that using either \autoref{eq:final-constraint-on-Om} or \autoref{eq:constraint-Omh008} as a prior only makes a negligible difference.

\section{Joint constraints with post-recombination and non-local observations}\label{sec:joint-analysis-with-other}
Several post-recombination observations are independent of or insensitive to early-universe physics; these are summarized in \citet{Lin_Mack_Hou_2020}. When combined with UCS, they can break the degeneracy in the cosmic background evolution and provide early-universe-physics insensitive constraints on $H_0$. Specifically, we will consider the cosmic chronometers, $\gamma$-ray optical depth, cosmic age and large-scale structure (LSS), respectively. Note that the joint constraints are not local measurements; they depend on a post-recombination cosmological model. The constraints on $H_0$ using UCS+non-local observations trade the dependence on early-universe physics with that on other astrophysical effects. Studies of how to mitigate and/or better account for astrophysical uncertainties in each observation are warranted. In this work, we include estimates of systematic uncertainties from the latest works; see discussions in each entry below. Potentially unaccounted for systematic errors would cause biases in their $H_0$ inferences, but the agreement between results from the different observations we consider suggest the consistency with current published estimates of systematic errors. LSS to some extent brings back the dependence on early-universe physics but still provides a sound-horizon insensitive constraint on $H_0$; see a later discussion.

\textbf{Cosmic chronometers (CC):}
The cosmic chronometer is a technique that directly measures the Hubble parameter $H(z)$ by measuring the differential ages of passively evolving galaxies at two nearby redshifts \citep{2016-Moresco-etal,2017-Moresco-Marulli-CC,2020-Moresco-Raul-Verde}. This technique does not depend on early-universe physics, but may suffer from various astrophysical systematic errors \citep{2020-Moresco-Raul-Verde,2017-Moresco-Marulli-CC}. Therefore, we provide various results taking into account additional systematic errors considered in \citet{2020-Moresco-Raul-Verde}. The results include analyses of the data from: 1) the current public release (current); 2) with uncorrelated systematic errors in the ``odd-one-out'' scenario (extra systematic); 3) with systematic errors in the ``odd-one-out'' scenario but conservatively include the full correlation of those systematic errors at different redshifts using (9) in \citet{2020-Moresco-Raul-Verde} (extra systematic, conservative). Note that the last case is a conservative estimation of extra systematic errors because systematic errors for data at different redshift bins are actually not quite correlated \citep{2020-Moresco-Raul-Verde}.

\textbf{The $\gamma$-ray optical depth:} High energy $\gamma$-ray photons interact with the diffuse extragalactic background lights and pair-create electrons and positrons \citep{Gould-Schreder-1996}. The rate of $\gamma$-ray photons being attenuated depends on the proper density of the extragalactic background light as well as the proper distance along the line of sight, and hence depends on the cosmic background evolution. Measuring the $\gamma$-ray optical depth has thus provided an independent constraint on the background evolution \citep{Ackermann-etal-science-2012,H.E.S.S-collaboration-2013,Biteau-Williams-2015,Dominguez-etal-2019}. The extragalactic background light is mainly produced by star formation; \citet{Dominguez-etal-2019} studied the systematic errors due to different models in the calculations of the density of extragalactic light. We obtain the MCMC chain of the analysis carried in \citet[private comminication]{Dominguez-etal-2019} and estimate the joint constraint of UCS+$\gamma$-ray by adding weights to each sampled ($H_0$, $\Omega_{\rm{m}}$) of the chain according to \autoref{eq:constraint-Omh008}. 

\textbf{Cosmic age:} Since most of the cosmic time is after recombination (in fact, the cosmic time after $z=100$ contributes about $99.9\%$ of the entire cosmic age), measuring ages of individual old stars, globular clusters and galaxies is a promising way to test early-universe resolutions to the Hubble tension \citep{Lin_Mack_Hou_2020,2021-Bernal-etal-new-cosmic-triangle,2021-Boylan-Kolchin-Weisz-redshift-z-relation}. With some prior on the time or redshift when the old stars formed, the age of the universe $t_{\rm{U}}$ can be inferred \citep{2019-Jimenez-etal-stellar-and-cosmic-age}. The age determination of stars and globular clusters is systematic-error dominated \citep{O_Malley-etal-2017,2017-Wagner-Kaiser-et-al,2019-Jimenez-etal-stellar-and-cosmic-age}. Recently, \citet{2021-Valcin-etal-cosmic-age} updated the age estimate of the oldest galactic globular clusters by reducing the systematic uncertainty due to the depth of the convection envelope, which is the most important known systematic error. They infer $t_{\rm{U}}=13.5\pm0.27$\,Gyr based on a spectroscopic metallicity determination and $t_{\rm{U}}=13.5\pm0.33$\,Gyr relying only on the clusters' color-magnitude diagrams.\footnote{While there is a limit on how early the actual formation time of the old stars can be, if they formed later, the inferred cosmic age would be higher and the inferred $H_0$ would be lower. } We adopt these cosmic age estimations and combine with UCS to provide early-universe insensitive constraints on $H_0$.

\begin{table*}[tbp]
\begin{ruledtabular}
\caption{Summary of the constraints on $H_0$.
The first subgroup contains the improved results using the method of this paper, namely, combining UCS with each of the listed early-universe-physics independent or insensitive observations. (See the text for detailed descriptions for each case.) They are all more consistent with Planck (the full standard analysis) or the TRGB-based local measurement from \citet{Freedman-etal-2020} than with the Cepheid-based local measurement.  For non-local determinations, we additionally include the results without using $\theta_{\rm{cmb}}$ in UCS (i.e. only using U.C. BAO and Pantheon as UCS). Comparing the results with and without $\theta_{\rm{cmb}}$, we can see the various extents to which non-local results improve their uncertainties and aggravate their tensions with the the Cepheid+SN Ia determination (R21). Strictly speaking, CMBlens+DES+BBN is not independent of early-universe physics, but depends on early-universe physics in a very different way from the methods that rely on the modeling of the sound horizon; see the text for a discussion.  Since all non-local results are consistent and independent with each other, for completeness we also provide the constraint on $H_0$ from jointly analyzing UCS+all late-time but non-local observations using a likelihood analysis. The UCS+all non-local result is in $2.4\sigma$ tension with R21. For comparison, in other entries we cite the constraints on $H_0$ from other measurements that are independent of early universe physics. (Most updated results from independent groups are cited.)
\label{tab:H0-constraints-summary}}
\begin{tabular}{p{0.55\linewidth}llcc}
 Methods  & \multicolumn{2}{c}{ $H_0$ (km/s/Mpc)} &  \multicolumn{2}{c}{$n$-$\sigma$ from R21} \\
\hline
\textbf{UCS+individual non-local observation} & Without $\theta_{\rm{cmb}}$ & \textbf{With $\theta_{\rm{cmb}}$}  & Without $\theta_{\rm{cmb}}$ & \textbf{With $\theta_{\rm{cmb}}$} \\  
~~~~ Cosmic Chronometers  &  \\
~~~~~~~~~~ Current public data \dotfill & $69.1\pm1.7$ &  $\bm{68.8\pm1.6}$ & 1.9$\sigma$ &  $\bm{2.1\sigma}$\\
~~~~~~~~~~ Extra systematic \dotfill & $69.4\pm2.3$ & $\bm{69.2\pm2.1}$ & 1.4$\sigma$ & $\bm{1.6\sigma}$\\
~~~~~~~~~~ Extra systematic, conservative \dotfill & $69.3\pm3.4$ & $\bm{68.9\pm3.3}$ & 1.1$\sigma$ & $\bm{1.2\sigma}$ \\
~~~~ $\gamma$-ray optical depth \dotfill & $66.2\pm3.5$ & $\bm{66.1\pm3.}4$ & 1.9$\sigma$  & $\bm{2.0\sigma}$\\ 
~~~~ Cosmic Age  \\
~~~~~~~~~~ $t_{\rm{U}}=13.5\pm0.27$\,Gyr\dotfill & $70.2\pm1.7$ & $\bm{69.8\pm1.5}$ & 1.4$\sigma$ & $\bm{1.7\sigma}$ \\
~~~~~~~~~~ $t_{\rm{U}}=13.5\pm0.33$\,Gyr\dotfill & $70.3\pm 2.1$ & $\bm{69.8\pm1.9}$  & 1.2$\sigma$ & $\bm{1.5\sigma}$ \\
~~~~ CMBlens+DES+BBN \dotfill & $68.8\pm2.4$ & $\bm{68.6\pm2.0}$  & 1.6$\sigma$ & $\bm{1.9\sigma}$\\
\textbf{UCS+joint non-local observations}\,$^{a}$\\ 
~~~~ All non-local observations \dotfill & $69.1\pm1.5$ &  $\bm{68.8\pm1.3}$  & 2.0$\sigma$ & $\bm{2.4\sigma}$\\
~~~~ Non-local observations without cosmic age\dotfill & $68.3\pm1.9$ & $\bm{68.1\pm1.6}$ & 2.1$\sigma$ & $\bm{2.5\sigma}$\\
~~~~ Non-local observations without LSS\dotfill & $69.1\pm1.6$ & $\bm{68.8\pm1.5}$ & 2.0$\sigma$ & $\bm{2.2\sigma}$\\
\hline
\textbf{Time-delay strong-lensing}\,$^{b}$ \\
\multicolumn{2}{l}{~~~~ TDCOSMO \citep{2020-Millon-etal} \dotfill} & $74.2\pm1.6$\\
\multicolumn{2}{l}{~~~~ TDCOSMO+SLACS  ~\citep{Birrer-2020} \dotfill} & $67.4^{+4.1}_{-3.2}$\\
\hline
\textbf{Local measurements}\,$^{c}$ \textbf{(distance ladder)} & \\
\multicolumn{2}{l}{~~~~ Cepheid+SN Ia \citep{2021-Riess-etal-H0} \dotfill} & $73.2 \pm 1.3$ \\
\multicolumn{2}{l}{~~~~ TRGB+SN Ia (a) \citep{Freedman-etal-2020} \dotfill} & $69.8\pm1.9$ \\
\multicolumn{2}{l}{~~~~ TRGB+SN Ia (b) \citep{Yuan-etal-2019-TRGB-local} \dotfill} & $72.4\pm2.0$ \\
\multicolumn{2}{l}{~~~~ TRGB+SN Ia (c) \citep{2020-Soltis-Casertano-Riess}\dotfill} & $72.1\pm2.0 $ \\
\multicolumn{2}{l}{~~~~ Mira+SN Ia \citep{2019-Huang-etal-Mira-H0} \dotfill} & $73.3\pm3.9$ \\
\multicolumn{2}{l}{~~~~ Cepheid+SBF+SN Ia \citep{Khetan-etal-2020} \dotfill} & $70.5\pm4.1$ \\
\multicolumn{2}{l}{~~~~ Cepheid/TRGB+SBF \citep{2021-Blakeslee-etal} \dotfill} & $73.3\pm2.5$ \\
\multicolumn{2}{l}{~~~~ Cepheid/TRGB+TFR \citep{2020-Kourkchi-etal,2020-Schombert-etal}  \dotfill} & $76.0\pm2.5$ \\ 
\multicolumn{2}{l}{~~~~ Cepheid/TRGB+SN II \citep{2020-Jaeger-etal} \dotfill} & $75.8^{+5.2}_{-4.9}$ \\
\hline
\textbf{Local measurements (non-distance ladder)}  \\
\multicolumn{2}{l}{~~~~ Megamaser Cosmology Project \citep{Pesce-etal-2020} \dotfill} & $73.9\pm3.0$\\
\multicolumn{2}{l}{~~~~ Standard siren multi-messenger \citep{2019-GW-BNS-update}  \dotfill} & $69^{+16}_{-8}$ \\
\end{tabular}
\begin{tablenotes}
\item $^{{a}}$ Conservative settings in cosmic chronometers and cosmic age are used.
\item $^{b}$ The time-delay strong-lensing technique is not a method of local measurement, but is insensitive to the underlying cosmological model. When the standard-$\Lambda$CDM is assumed, its determination of $H_0$ is insensitive to $\Omega_{\rm{m}}$. Compared to the TDCOSMO entry, the result in TDCOSMO+SLACS relaxes the strong assumption of the lens mass density profile, considers correlations of the modeling between lenses, and adds imaging and spectroscopic data of other lenses. 
\item $^{c}$ Local measurements based on the distance ladder technique are represented as primary+secondary distance indicators. SBF stands for Surface Brightness Fluctuation, Mira for Mira variables, TFR for Tully Fisher Relation and SN II for Type II Supernova. Note that not all local measurements are independent, since some of them use the same type of primary or secondary distance indicator. It is also worth pointing out that the zero point of the TRGB used in Cepheid/TRGB+TFR \citep{2020-Kourkchi-etal,2020-Schombert-etal}, Cepheid/TRGB+SN II \citep{2020-Jaeger-etal} and Cepheid/TRGB+SBF \citep{2021-Blakeslee-etal} are somewhat different from that in TRGB+SN Ia (a) \citep{Freedman-etal-2020}.
\end{tablenotes}
\end{ruledtabular}
\end{table*}

\textbf{The large-scale structure with big bang nucleosynthesis (LSS+BBN):}
Earlier, it was pointed out that the combination of several current LSS observations can provide an independent constraint on $H_0$ \citep{WL2017b}. But some observations used, such as the redshift space distortion (with BAO included), mix the dependence on the sound horizon scale and thus are not independent of early-universe physics. Later, \citet{2020-Baxter-Sherwin,2020-Philcox-Sherwin-Farren-Baxter,2020-Pogosian-Zhao-Jedamzik} showed that subsets of LSS observations with uncalibrated SNIa and/or uncalibrated BAO can provide a sound-horizon independent constraint on $H_0$. The idea is that the shape of the matter power spectrum is sensitive to the horizon scale at matter-radiation equality which depends on matter-to-radiation energy ratio today parameterized by $\Omega_{\rm{m}}h^2$ \citep{2020-Philcox-Sherwin-Farren-Baxter}. Since only the distance normalized by $1/H_0$ can be measured, LSS can put a constraint on $\Omega_{\rm{m}}h$ and, with our UCS constraint (mainly) on $\Omega_{\rm{m}}$, can break the degeneracy in the $H_0$-$\Omega_{\rm{m}}$ plane to provide a constraint on $H_0$. In this work, we use the CMB lensing from Planck and the $3\times2$ correlation functions from DES (also used in \citet{2020-Pogosian-Zhao-Jedamzik}), as the two together can probe the matter power spectrum on a wide range of scales including resolving the peak of the matter power spectrum \citep{2020-Baxter-Sherwin,2020-Pogosian-Zhao-Jedamzik}. The matter power spectrum has a weak dependence on $\Omega_{\rm{b}}h^2$ and we adopt a prior of $\Omega_{\rm{b}}h^2=0.0222\pm0.005$ obtained from BBN \citep{Cooke-etal-2018}.  Note that, unlike the other non-local observations, combining LSS to some extent raises the sensitivity of UCS to early-universe physics. However, LSS depends on early-universe physics in a very different way from that of the sound horizon scale, which is the most important quantity for the CMB inference of $H_0$. As mentioned earlier, \citet{2020-Baxter-Sherwin,2020-Philcox-Sherwin-Farren-Baxter,2020-Pogosian-Zhao-Jedamzik} have shown the insensitivity of LSS (as well as the BBN constraint on $\Omega_{\rm{b}}h^2$) to sound-horizon physics. Relating to this, the introduction of early-universe physics that decreases the sound horizon tends to worsen the $\sigma_8$ tension \citep{2020-Jedamzik-Pogosian-Zhao}.

We summarize the results of the joint analyses with each of the above mentioned observations and UCS in Table \ref{tab:H0-constraints-summary}. All constraints prefer a lower value of $H_0$ that is more consistent with Planck and the Tip of the Red Giant Branch (TRGB)-based local measurement\footnote{It is however worth pointing out that the TRGB-based result remains controversial; see \citet{Freedman-etal-2019,Yuan-etal-2019-TRGB-local,Freedman-etal-2020,2020-Soltis-Casertano-Riess}.} from \citet{Freedman-etal-2020} than with the Cepheid-based local measurement. In particular, the fact that our result from UCS+CMBlens+DES+BBN prefers a lower value of $H_0$ is consistent with the finding that combining Planck and LSS disfavors early dark energy \citep{Hill-etal-2020,2020-Ivanov-McDonough-Hill-early-DE-lss}.

Since all non-local results are consistent with and independent of each other, for completeness we jointly analyze UCS+all late-time but non-local observations. In doing so, we adopt conservative error estimates on cosmic chronometers and cosmic age. The joint constraint on $H_0$ is in a $2.4$-$\sigma$ tension with the Cepheid+SN Ia determination. In addition, we provide one joint result excluding the cosmic age due to its prior on the formation time of the globular cluster and some other potentially unaccounted-for systematic errors, and one excluding LSS so that the remaining constraints are highly insensitive to early-universe physics. These two weaker joint results are also in tension with the Cepheid+SN Ia determination with $>2\sigma$.

For comparison, we also list in Table \ref{tab:H0-constraints-summary} other measurements of $H_0$ including the results from the time-delay strong-lensing technique (TDCOSMO) \citep{Birrer-2020} and local measurements. Although with larger uncertainties than that of the Planck 2018 inference of $H_0$, the post-recombination but non-local constraints on $H_0$ appear to have some tension with  (and are all consistently lower than) most of the local measurements, except for the TRGB+SN Ia result reported in \citep{Freedman-etal-2020}. Incidentally, it is worth noting that not all local measurements are totally independent from each other, because some of them share the same type of primary or secondary distance indicator; see Table \ref{tab:H0-constraints-summary}. 

The tension we find here may be due to either post-recombination new physics, unknown systematic errors in some local measurements or unknown systematic errors in other non-local observations, but cannot be resolved by introducing early-universe physics. Regarding this, the fact that non-local results are independent in terms of statistics and methodology but still consistent with each other disfavors the explanation of systematic errors in non-local observations. Our results challenge early-universe-physics resolutions to the Hubble tension, which is in agreement with \citet{2020-Jedamzik-Pogosian-Zhao} who come to a similar conclusion in a very different way.

\section{Discussion}\label{sec:discussion}
In our analyses, we have assumed a flat-$\Lambda$CDM universe after recombination since we are interested in constraining the cosmic background without a strong assumption of the standard cosmological model at early times. But our method can also be generalized to allow a nonstandard cosmological model after recombination. This can serve as a strong and early-universe-physics insensitive test to late-time models, such as voids \citep{2020-Ding-Nakama-Wang-VOID,2019-Shanks-etal-local-voids1,2019-Shanks-etal-local-voids2} and modified gravity/dark energy models \citep{Pan-etal-2019-interacting-DS,Belgacem-etal-2018-nonlocal-MG,2018-Park-Ratra,2021-Yang-Pan-DiValentino-Mena-Melchiorri,2020-Shimon-Weyl-invariant-gravity,2020-Gonzalez-Hertzberg-Rompineve}. The advantages of this test are: 1) the constraint on late-time models will be free from biases due to any possible early-universe nonstandard physics or systematic errors in the CMB data that affect the amplitude or the shape of the power spectrum; 2) while releasing the assumption of the standard cosmological model before recombination, the constraining power on the late-time background has been shown to be largely maintained. 

To use UCS as an early-universe-physics insensitive test on late-time cosmological models, all that is needed is to modify the normalized comoving distance accordingly (i.e., \autoref{eq:comoving-distance-in-H0-flat}). Note that one is able to constrain late-time cosmological models without constraining the value of $H_0$, because late-time models predict the relative change of $H(z)$ as a function of redshift instead of predicting its absolute value. UCS is a clear and strong probe of the relative change of $H(z)$. Therefore, when constraining late-time nonstandard cosmological models, it is neither necessary nor recommended to adopt a prior of $H_0$ obtained from the Cepheid-based local measurement. It is also unnecessary to combine with the observations listed in Sec.\,\ref{sec:joint-analysis-with-other}, as the test uses a minimal set of observations and still remains strong, although doing so allows us to gain some additional constraining power.

We also note that \citet{2015-Aubourg-BOSS-collaboration} has used the idea of treating the measurement of the CMB angular scale as an extra data point for BAO.\footnote{We thank Matias Zaldarriaga for bringing the reference \citep{2015-Aubourg-BOSS-collaboration} to our attention.} In their method, they assume the Standard Model expression of $r_d$ as a function of $\Omega_{\rm{m}} h^2$ and $\Omega_{\rm{b}} h^2$ but drastically increase the prior ranges for these two parameters. Effectively, $r_d$ becomes a free parameter because $\Omega_{\rm{b}}h^2$ is unconstrained. In our work, we directly leave $r_d$ as a free parameter and only model its difference from $r_*$. For the purpose of our paper, namely to disentangle the interference of the early universe physics, the latter approach is more advantageous, because e.g.~it removes the dependence on BBN which (when applied) can determine the parameter $\Omega_{\rm{b}}h^2$ and thus constrain $r_{\rm{d}}$ (and $h$) by the Standard Model in the former approach.

\section{Summary and conclusion}\label{se:summary-and-conclusion}
Constraining the Hubble constant in an early-universe-physics independent or insensitive way is important, because this can check whether the current tension between Planck and the Cepheid-based local measurement is due to some undiscovered early-universe physics. In this work, we have introduced a method that straightforwardly analyzes $\theta_{\rm{cmb}}$ in an early-universe-physics insensitive way.
We jointly analyzed the angular size of CMB acoustic peaks, baryon acoustic oscillation and Type Ia supernovae by treating their absolute scales or luminosity as free parameters. 

These uncalibrated standard rulers and candles, measured across cosmic time from recombination to today, provide a strong constraint on the cosmic background evolution insensitive to early-universe physics. When a flat $\Lambda$CDM is assumed after recombination, UCS mainly constrains the matter density parameter today, but the exact constraint exhibits a small positive degeneracy in the $H_0$-$\Omega_{\rm{m}}$ plane.

The method of UCS can also be generalized and used to test post-recombination nonstandard cosmological models in a way that is insensitive to modifications of early-universe physics or systematic errors that affect the amplitude of the CMB power spectra at different scales, yet can still be powerful.

To analyze the problem of the Hubble tension in a way that is insensitive to pre-recombination physics, we combined UCS with other early-universe-physics independent or insensitive, but non-local, observations to break the degeneracy in the $H_0$-$\Omega_{\rm{m}}$ plane and obtained several constraints on $H_0$. The combination with LSS mildly raises the dependence to early-universe physics, but still insensitive to the sound-horizon physics as discussed in \citet{2020-Baxter-Sherwin,2020-Philcox-Sherwin-Farren-Baxter,2020-Pogosian-Zhao-Jedamzik,2020-Jedamzik-Pogosian-Zhao} and its result on $H_0$ is consistent with other non-local determinations. All non-local constraints are more consistent with the Planck and a TRGB-based local measurement than with the Cepheid-based local measurements (see Table.~\ref{tab:H0-constraints-summary}). 

In the literature, the Hubble tension is usually attributed to some discrepancy between the early (pre-recombination) and late (post-recombination) universe. Our analyses provide some new insights into this statement. Hubble-constant measurements of the late universe include both non-local and local measurements. Our results suggest that the non-local ones are still consistent with the CMB result, and the tension has more to do with a tension between the non-local (including both pre- and post-recombination) measurements and most of the local measurements.

This discrepancy may be due to new physics in the post-recombination epoch, or may disappear after future improvement of systematic errors in some local measurements or all non-local observations. Indeed, since the $H_0$ determinations using UCS+non-local observations significantly reduce the dependence on early-universe physics, it suggests an increased focus on astrophysical effects. While we have adopted the latest treatments in each non-local observation, more studies are warranted to reduce and better treat the systematic uncertainties in each observation. Fortunately, the fact that all non-local probes are independent in methodologies and that their constraints on $H_0$ (jointly with UCS) are consistent with each other suggests an absence of major unaccounted-for systematic errors in those non-local observations. In any case, this tension between the local determination and UCS+non-local results will not be resolved by introducing nonstandard physics in the pre-recombination early universe.

Identifying the cause of the reported inconsistencies---being either beyond-the-standard-model physics or unaccounted-for/underestimated systematic errors---has become and will continue to be an important topic in cosmology; see, e.g., \citet{Lin_Mack_Hou_2020,2014Efstathiou-Hubble,2020-Yao-Shan-Zhang-Kneib,2019-Garcia-Ishak-Fox-Lin,2018Yu-Ratra-Wang,WL2017b,2020-Yao-Shan-Zhang-Kneib,2016-Hubble-reconcile,2018-Gomez-Valent-Sola,Vagnozzi-2019,Yang-eta-2019-many-w}. Reanalyzing data by relaxing strong assumptions is a useful way to check those assumptions. This philosophy has been used to check the validity of other astrophysical assumptions as well, such as in the strong-lensing time-delay technique of determining the Hubble constant \citep{Birrer-2020,2019Gomer-etal,Blum-etal-2020}. In the future, galaxy surveys such as the Dark Energy Spectroscopic Instrument (DESI) will provide measurement of BAO with unprecedented precision  \citep{2016-DESI-white}. The Vera C.~Rubin Observatory will discover many more Type Ia supernovae \citep{2012-LSST-white-paper}. Measuring BAO from 21cm anisotropy will fill in standard rulers during cosmic dawn \citep{2019-Munoz-VAO}. In the meantime, the techniques using cosmic chronometers, $\gamma$-ray optical depth and large-scale structure will be improved, which will help provide more precise and robust measurement of $H_0$ without strongly assuming early-universe physics. All these data will further reduce the errors and make the method of UCS more powerful. With these, we shall soon be able to narrow down the cause of the Hubble tension.

\begin{acknowledgements}
We thank Gongjun Choi for helpful discussions, and Eoin Colgain, Michele Moresco, Levon Pogosian, Matias Zaldarriaga, and an anonymous referee for helpful feedback and comments. 
\end{acknowledgements}

\appendix

\section{Data and Likelihoods}\label{sec:data-and-likelihoods}
\subsection{Notations and background evolution}\label{sec:notation-and-background}
Here we discuss in detail the framework and the UCS data we used in this work. We normalize the (transverse) comoving distance $d_{\rm{M}}$ by the Hubble distance ($1/H_0$),
\begin{equation}\label{eq:comoving-dist-DM}
d_{\rm{M}}(z) = \frac{1}{H_0}f_{\rm M}(z; \,\Omega_{\rm{m}},\cdots)\,,  
\end{equation}
with $f_{\rm M}(z; \,\Omega_{\rm{m}},\cdots)$ dependent on a cosmological model.  We assume the standard spatially flat $\Lambda$CDM model after recombination and,
\begin{eqnarray}
&f_{\rm M}(z;\,\Omega_{\rm{m}},\cdots) = \int_{0}^z\frac{dz'}{E(z')}\,,\label{eq:comoving-distance-in-H0-flat}\\
&E(z)=\sqrt{\Omega_\Lambda+\Omega_{\rm{m}}(1+z)^3+\Omega_{\rm{r}}(1+z)^4+\sum\limits_i\Omega_{\nu_i}e_i(z)}\,,\label{eq:Eofz}
\end{eqnarray}
with the constraint $1=\Omega_{\rm{m}}+\Omega_\Lambda+\Omega_{\rm{r}}+\sum\limits_i\Omega_{\nu_i}$. The subscript `m' stands for matter, `$\Lambda$' for the cosmological constant, `r' for radiation (photon and massless neutrino), and `$\nu_i$' for massive neutrinos where $i$ runs through the neutrino species. The function $e_i(z)$ represents the redshift dependence of the density fraction evolution of the massive neutrinos and can be well approximated by \citep{2021-Lin-DM} 
\begin{equation}\label{eq:evolution-nu}
    e_i(z)=\frac{1}{a^4}\left(\frac{a^n+a_{{\rm{T}},i}^n}{1+a_{{\rm{T}},i}^n}\right)^{\frac{1}{n}}\,,
\end{equation}
where $a=\frac{1}{1+z}$ is the scale factor, $n=1.8367$, $a_{{\rm{T}},i}=3.1515\frac{T_\nu^0}{m_{\nu_i}}$, $T_\nu^0=1.676\times10^{-4}$\,eV and $m_{\nu_i}$ is the $i$th neutrino mass. The radiation and neutrino terms only have some small contribution to $f_{\rm{M}}$ at redshifts close to recombination but we include them for completeness. In our standard analysis, we consider one massless and two massive neutrinos in the normal hierarchy with constraints on the difference of the mass-squared $\Delta m^2_{12}=7.9\times10^{-5}\,{\rm eV}^2$ and $\Delta m^2_{13}=2.2\times10^{-3}\,{\rm eV}^2$  \citep{2004-Maltoni-Schwetz-nu-masses}. With the measured CMB temperature today and the assumption of thermal neutrino relics, we have $\Omega_{\rm{r}}=\Omega_{\gamma}+\Omega_{\nu}^{\rm{massless}}=3.0337\,h^{-2}\times10^{-5}$ and $\Omega_{\nu_i}=\frac{ m_{\nu_i}}{93.14\,h^2\, {\rm eV}}$ where $h\equiv\frac{H_0}{100\,\rm{km/s/Mpc}}$ \citep{2012-Lesgourgues-Pastor-nu-mass}. While we have ignored the uncertainties of the mass-squared difference and the mass of the lightest neutrino, we found that our results are insensitive to them for a reasonable range of the lowest neutrino mass for both hierarchy schemes. This is because the radiation and neutrino terms only have small effects on $f_{\rm{M}}$ at high redshifts close to recombination. For a reasonable range of the mass of the lightest neutrino, all neutrinos were relativistic before $z\sim100$ when they have (small) effects on $f_{\rm{M}}$. In fact the effects of the radiation and the neutrinos terms are so small that our constraints on $\Omega_{\rm{m}}$ are essentially unchanged even if we completely ignore these two terms.

In our standard treatment, $f_{\rm{M}}$ mainly depends on $\Omega_{\rm{m}}$. From our adopted normalization [i.e., \autoref{eq:comoving-dist-DM}] we shall see more clearly the degeneracy of $H_0$ with the scale of the sound horizon ($r_*$ or $r_{\rm{d}}$) and with the absolution magnitude of SNe Ia, $M_0$. However, it is worth pointing out that $f_{\rm{M}}$ still weakly depends on $h$ via the radiation and neutrino terms, especially at higher redshifts\footnote{In other words, if we can measure UCS extremely well, $H_0$ can be inferred even without any calibration to cosmic standards!}. As a consequence, our constraints will not be totally uninformative in $H_0$; see Sec.\,\ref{sec:constraint-on-H0-Om}.

Note that in this work, we have assumed the standard $\Lambda$CDM for cosmic time after recombination, but our analyses can be generalized and used to test post-recombination nonstandard models where \autoref{eq:comoving-distance-in-H0-flat} is replaced by the corresponding normalized comoving distance. Such a test of post-recombination nonstandard models will have the advantage of being robust against any potentially unknown early-universe physics while retaining a strong constraining power.

\subsection{\label{sec:BAO}The late-time Baryon Acoustic Oscillation}\label{sec:appendix-bao}
The late-time BAO measurements are compressed information obtained from two-point correlation functions of tracers of the underlying matter fluctuation. With $r_{\rm{d}}$ the sound horizon scale at the end of the drag epoch (at redshift $z_{\rm{d}}$), measurements can be given by a pair of quantities which are the angular size $\theta_{\rm{d}}=\frac{r_{\rm{d}}H_0}{f_{\rm{M}}(z)}$ and the reshift span $\Delta z_{r_{\rm{d}}}= r_{\rm{d}}H_0E(z)$ of the baryon acoustic sound horizon when placed perpendicular to and along the light of sight. Alternatively, the set $\big[\frac{r_{\rm{d}}H_0}{f_{\rm{V}}(z)},~F_{\rm{AP}}(z)=\frac{\Delta z_{r_{\rm{d}}}}{\theta_{\rm{d}}}=E(z)f_{\rm{M}}(z)\big]$ will be given or can be constructed from $(\theta_{\rm{d}},~\Delta z_{r_{\rm{d}}})$. Here
\begin{equation}\label{eq:fV}
f_{\rm{V}}(z)\equiv\Big[\frac{z}{E(z)}\big[f_{\rm M}(z)\big]^2\Big]^{1/3}\,, 
\end{equation}
which can be interpreted as the cubic root of the differential comoving volume (w.r.t. the solid angle and logarithmic redshift) normalized by $(1/H_0)^3$. The quantity $\frac{r_{\rm{d}}H_0}{f_{\rm{V}}(z)}$ can be interpreted in the following way. Imagine a simple cubic lattice formed in the universe with the comoving length of each dimension equal to the size of the drag sound horizon. The comoving volume element of each lattice cell is $r_{\rm{d}}^{~3}$. Then, $\big(\frac{r_{\rm{d}}H_0}{f_{\rm{V}}(z)}\big)^3$ is (the inverse of) the number of those volume elements seen per solid angle per logarithmic redshift. The quantity $F_{\rm{AP}}(z)$ is the ratio between $\Delta z_{r_{\rm d}}$ and $\theta_{\rm{d}}$, which is useful as it contains no dependence of $\Delta rH_0$.  

It is worth pointing out that $r_{\rm{d}}$ always goes with $H_0$ in those BAO measurements, while $f_{\rm{M}}$ and $f_{\rm{V}}$ are nearly independent of $H_0$. So, when $r_{\rm{d}}$ is treated as a free parameter, it is degenerate with $H_0$; also see \citet{Aylor-etal-2019}. We therefore treat $r_{\rm{d}}H_0$ as one free parameter when analyzing the BAO data. 

In Table \ref{tab:BAO-measurement} we summarize some BAO measurements obtained or inferred from (in ascending order with effective redshift) the 6dF galaxy survey (6dF) \citep{Beutler-etal-2011}, SDSS DR7 `main galaxy sample' (MGS) \citep{Ross-etal-2015}, SDSS DR12 galaxy clustering \citep{Alam-etal-2017}, SDSS DR16 luminous red galaxy and emission line galaxy samples (LRG \& ELF) \citep{2020-Wang-etc}, and eBOSS DR16 quasars (QSO) and Lyman-$\alpha$ forest (Ly-$\alpha$) auto and cross correlations \citep{Blomqvist-etal-2019}. There are correlations between $\theta_{\rm{d}}$ and $\Delta z_{r_{\rm{d}}}$, which are shown in the last column of Table \ref{tab:BAO-measurement}. In particular, there are correlations among the three DR12 GC BAO measurements with
\begin{equation}\label{eq:correlation-consensus}
\bm{\rho}_{_{\rm{DR12 GC}}}(\bm{\theta_{\rm{d}}},\bm{\Delta z_{r_{\rm{d}}}})=
   \begin{pmatrix}
       1 & -0.084 &  0.47 & -0.042 &  0.17 & -0.014 \\
       -0.084 &  1 & -0.012 &  0.41 &  0.012 &  0.16 \\
       0.46 & -0.012 &  1 & -0.027 &  0.50 & -0.022 \\
       -0.042 &  0.41 & -0.027 &  1 -0.0085 & 0.47 \\
       0.17 &  0.012 &  0.50 & -0.0085  &  1 & 0.050 \\
       -0.014 &  0.16 & -0.022 & 0.47 &  0.050 & 1
    \end{pmatrix}\,.
\end{equation}

The uncalibrated BAO likelihood reads
\begin{equation}\label{eq:bao-likelihood}
    \ln \mathcal{L}_{\rm{\textsc{u.c.\,bao}}}=\sum\limits_i-\frac{1}{2}(\bm{\Theta_i}-\bm{Q_i})^T\bm{C_i}^{-1}(\bm{\Theta_i}-\bm{Q_i})+\rm{Const.,}
\end{equation}
where $\bm{\Theta_i}(\Omega_{\rm{m}},r_{\rm{d}}H_0)$ and $\bm{Q_i}$ are predictions and measurements of $(\theta_{\rm{d}},~\Delta z_{r_{\rm{d}}})$ or $\big(\frac{r_{\rm{d}}H_0}{f_{\rm{V}}(z)},~F_{\rm{AP}}(z)\big)$, respectively, and $\bm{C_i}$ is the covariance matrix and the index $i$ denotes different BAO measurements.

\begin{table*}[htbp]
\caption{BAO measurements obtained or inferred from \citet{2020-Wang-etc,Blomqvist-etal-2019,Alam-etal-2017,Ross-etal-2015,Beutler-etal-2011}.}
    \label{tab:BAO-measurement}
    \footnotesize
    \begin{ruledtabular}
    \begin{tabular}{lcc|cc|ccc}
        Labels & Refs. & $z_{\rm{eff}}$  & $\frac{r_{\rm{d}}H_0}{f_{\rm{V}}(z)}$ & $F_{\rm{AP}}$ &  $\theta_{\rm{d}}\equiv\frac{r_{\rm{d}}H_0}{f_{\rm{M}}(z)}$ & $\Delta z\equiv r_{\rm{d}}H_0\times E(z)$  & $\rho(\theta_{\rm{d}},\Delta z_{r_{\rm{d}}})$\\
        \hline
        6df &\citep{Beutler-etal-2011} & 0.106 & $0.336\pm0.015$ & N/A & N/A & N/A & N/A \\
        MGS &\citep{Ross-etal-2015} & 0.15 & $0.224\pm0.0084$ & N/A & N/A &  N/A & N/A\\
        \multirow[t]{3}{*}{DR12 GC} & \multirow[t]{3}{*}{\citep{Alam-etal-2017}} & 0.38 & $0.1002\pm0.0011$ & $0.410\pm0.016$ & $0.0977\pm0.0016$  & $0.0400\pm0.0012$ & \autoref{eq:correlation-consensus} \\
        & & 0.51 & $0.07893\pm0.00080$ & $0.599\pm0.021$ & $0.0748\pm0.0011$ & $0.0448\pm0.0011$ & \autoref{eq:correlation-consensus}\\
        & & 0.68 & $0.06898\pm0.00071$ & $0.761\pm0.027$ & $0.0641\pm0.0010$ &  $0.0488\pm0.0012$ & \autoref{eq:correlation-consensus} \\
        LRG \& LEG & \citep{2020-Wang-etc} & 0.77 & $0.05707\pm0.00090$ & $0.960\pm0.035$ & $0.0530\pm0.0011$ & $0.0509\pm0.0015$ & $8.02\times10^{-4}$ \\ 
        Quasar & \citep{Blomqvist-etal-2019} & 1.52 & $0.0383\pm0.0017$ & N/A & N/A & N/A & N/A \\
        Ly$\alpha$-Quasar & \citep{Blomqvist-etal-2019} & 2.35 & $0.03256\pm0.00065$ & $4.13\pm0.20$ & $0.02698\pm0.0009$ & $0.1115\pm0.0028$ & $0.645$
    \end{tabular}
    \end{ruledtabular}
\end{table*}

\subsection{The angular size of the acoustic scale from CMB}\label{sec:appendix-CMB-theta}
The photon-baryon plasma perturbation also manifests as acoustic oscillations in the CMB temperature and polarization angular power spectra. These oscillations correspond to a sharply-defined angular scale ($\theta_{\rm{cmb}}$), which has been robustly and precisely measured and is almost independent of the underlying cosmological model \citep{2018-Planck-cosmo-params}. It reads $\theta_{\rm{cmb}}=r_*/d_{\rm{M}}^{\rm{rec}}$, where $r_*$ is the comoving sound horizon at recombination and $d_{\rm{M}}^{\rm{rec}}$ is the comoving distance at recombination. Plugging in \autoref{eq:comoving-dist-DM} we have $\theta_{\rm{cmb}}=\frac{r_*H_0}{f_{M}(z_*)}$. So $\theta_{\rm{cmb}}$ mainly depends on $r_*H_0$ and $\Omega_{\rm{m}}$. It weakly depends on $h$ via the radiation and neutrino terms in $E(z)$ within $f_{\rm{M}}(z)$; see Appendix \ref{sec:notation-and-background}. It also depends on $z_*$. We first assume a prior on $z_*$ obtained from the Planck baseline constraint based on the standard $\Lambda$CDM model \citep{2018-Planck-cosmo-params} or from ACT+WMAP \citep{2020-Aiola-etal-ACT}. But since $f_{\rm{M}}(z\rightarrow\infty)\rightarrow{\rm{constant}}$, corresponding to the particle horizon normalized by the Hubble distance, and $z_*$ is sufficiently large, the dependence of $\theta_{\rm{cmb}}$ on $z_*$ is very weak. Therefore, our results would be barely changed for a reasonable range of $z_*$ and we verify this in Sec.\,\ref{sec:robustness-of-the-result}.

The $\theta_{\rm{cmb}}$ likelihood reads,
\begin{equation}\label{eq:cmb-theta-likelihood}
   \ln \mathcal{L}_{\theta_{\rm{cmb}}}=-\frac{1}{2}(\bm{\theta_{\rm{p}}}-\bm{\theta_{\rm{o}}})^T\bm{C}^{-1}(\bm{\theta_{\rm{p}}}-\bm{\theta_{\rm{o}}})+\rm{Const.,}
\end{equation}
where $\bm{\theta}=(\theta_{\rm{cmb}},z_*)$ and the subscripts `o' and `p' denote `observation' and `prediction', respectively. Here $z_*$ only serves as a nuisance parameter, and, since the derived constraint on $\theta_{\rm{cmb}}$ correlates with $z_*$ for both Planck and ACT+WMAP, we have included their correlation in the $\theta_{\rm{cmb}}$ likelihood \autoref{eq:cmb-theta-likelihood}. Thus, the parameters contained in the $\theta_{\rm{cmb}}$ likelihood are $r_*H_0$, $\Omega_{\rm{m}}$ and $h$ (a very weak dependence).

\subsection{Linking the two baryon acoustic sound horizons}\label{sec:link-between-bao-cmb}
The two sound horizon scales, $r_*$ at recombination and $r_{\rm{d}}$ at the end of the drag epoch, are closely related to each other. They are given by
\begin{align}
    r_*H_0&=\int^\infty_{z_*}\frac{c_{\rm{s}}(z)}{E(z)}dz\,,\label{eq:rsH0}\\
    r_{\rm{d}}H_0&=\int^\infty_{z_{\rm{d}}}\frac{c_{\rm{s}}(z)}{E(z)}dz\,,\label{eq:rdH0}
\end{align}
where $c_{\rm{s}}$ is the sound speed. A number of early-universe nonstandard models/considerations have been proposed to make both $r_*$ and $r_{\rm{d}}$ smaller to mitigate the Hubble tension between Planck and the Cepheid-based local measurement, e.g., \citet{Poulin-etal-2019-EDE,Kreisch-etal-2019-self-interacting-neutrinos,Pogosian-Jedamzik2020}. But those models (in fact, almost all early-universe nonstandard models) shift the two horizons by approximately the same constant, so their difference remains intact and small. (See discussions in Sec.\,\ref{sec:robustness-of-the-result} for the details and possible exceptions.) This permits us to jointly analyze the uncalibrated BAO and the $\theta_{\rm{cmb}}$ likelihoods by treating $r_*H_0$ as a free parameter and adopting some reasonable and model insensitive assumptions about the calculation of the difference between $r_*H_0$ and $r_{\rm{d}}H_0$.

To do this, we take the difference between \autoref{eq:rdH0} and \autoref{eq:rsH0},
\begin{equation}\label{eq:difference-rH0-app}
    \Delta rH_0\equiv(r_{\rm{d}}-r_*)H_0=\int^{z_*}_{z_d}\frac{c_{\rm{s}}(z)}{E(z)}dz\,.
\end{equation}
In our standard analysis, we first assume that $\Delta rH_0$ can be modeled in the same way as in the standard $\Lambda$CDM model, so that we have the sound speed given by
\begin{equation}\label{eq:sound-speed}
    c_{\rm{s}}(z;\,\Omega_{\rm{b}}h^2)=\frac{1}{\sqrt{3\big(1+\frac{1}{1+z}\frac{3\,\Omega_{\rm{b}}h^2}{4\,\Omega_{\gamma}h^2}\big)}}\,,
\end{equation}
and $E(z)$ given by \autoref{eq:Eofz}. With $\Omega_{\rm{\gamma}}h^2=2.47\times10^{-5}$, $c_{\rm{s}}(z)$ is determined if $\Omega_{\rm{b}}h^2$ is given. In the standard analysis, we assume a BBN prior on $\Omega_{\rm{b}}h^2=0.0222\pm0.0005$. In the $\theta_{\rm{cmb}}$ likelihood, we already assume a prior on $z_*$ obtained from Planck or ACT+WMAP, so the only thing needed to model $\Delta rH_0$ is the redshift difference between $z_*$ and $z_{\rm{d}}$. In the standard analysis, we also assume a prior on $\Delta z_{\rm{s}} \equiv z_*-z_{\rm{d}}$ obtained from Planck ($\Delta z_{\rm{s}}=30.26\pm0.57$) or ACT+WMAP ($\Delta z_{\rm{s}}=29.95\pm0.75$). Then, the joint $\theta_{\rm{cmb}}$ and uncalibrated BAO likelihood, i.e., $\ln \mathcal{L}_{\theta_{\rm{cmb}}}+\ln \mathcal{L}_{\rm{\textsc{u.c.\,bao}}}$, contain parameters $r_*H_0$, $\Omega_{\rm{m}}$ and $h$ (a very weak dependence), with some assumed priors on $z_*$, $\Delta z_{\rm{s}}$ and $\Omega_{\rm{b}}h^2$.

\subsection{Type-Ia Supernovae and a fast algorithm to compute the SN Ia likelihood}\label{sec:appendix-sn}
In an FLRW universe the luminosity distance (observed in the heliocentric frame) is related to the comoving distance by $d_{\rm L}=(1+z_{\rm{hel}})d_{\rm{M}}(z_{\rm{cmb}})$ (see, e. g., Eq. 2 in \citet{2019-Kessler-etal}) and we have
\begin{equation}\label{eq:magnitude-redshift}
    m = 5\log_{10}\big[(1+z_{\rm{hel}})f_{\rm{M}}(z_{\rm{cmb}}) \big]+\mathcal{M}\,,
\end{equation}
where $z_{\rm{hel}}$ and $z_{\rm{cmb}}$ are the redshifts of SNe Ia observed in the heliocentric frame and in the CMB frame, $m$ is the apparent magnitude of the SN Ia, and $\mathcal{M}$ is defined as
\begin{equation}\label{eq:C-H-M}
    \mathcal{M}\equiv M_0 -5\log_{10}(10\,{\rm{pc}}\times H_0)\,,
\end{equation}
with $M_0$ the absolute magnitude of SNe Ia at some reference stretch and color. We treat $\mathcal{M}$ as a free parameter in the SN Ia likelihood analysis. The SN Ia likelihood reads,
\begin{equation}\label{eq:sn-likelihood}
 \ln\mathcal{L}_{\rm{SN}}=
   -\frac{1}{2}(\bm{m_{\rm{o}}}-\bm{m_{\rm{p}}})^{\rm{T}}\times(\bm{C_{\rm{stat}}}+\bm{C_{\rm{sys}}})^{-1} 
  \times(\bm{m_{\rm{o}}}-\bm{m_{\rm{p}}})+\rm{Const.}    \,,
\end{equation}
where $\bm{m_{\rm{o}}}$ and $\bm{m_{\rm{p}}}(\Omega_{\rm{m}},\mathcal{M})$ are the observed and predicted apparent magnitudes, $\bm{C_{\rm{stat}}}$ and $\bm{C_{\rm{sys}}}$ are the statistic and systematic covariance matrices. In our standard analyses, the SN Ia likelihood has two free parameters, $\Omega_{\rm{m}}$ and $\mathcal{M}$. When analyzing the SN Ia data alone, we fixed $h=0.7$ for the radiation and neutrino terms. Since the relevant redshifts are low and these terms are very small, ignoring them leads to almost the same constraint on $\Omega_{\rm{m}}$ and $\mathcal{M}$. In this work, we use the Pantheon compilation of SNe Ia data \citep{Scolnic-etal-2018-Pantheon}. 

\begin{table*}[tbp]
    \centering
    \caption{The list of coefficients for one, two and three terms to be included in Eq.\,\eqref{eq:fM-i-Runge-kuta}. The two-term approximation is what we use with considerations of both numerical accuracy and speed.}
    \label{tab:b-c-coefficients}
    \begin{ruledtabular}
    \begin{tabular}{llll}
         & One term & Two Terms & Three terms \\
         \hline
        Coefficients & $c_1=1$, $b_1=\frac{1}{2}$ &  \pbox{20cm}{$c_1=\frac{1}{2}$, $b_1=\frac{1}{3-\sqrt{3}}$\\
        $c_2=\frac{1}{2}$, $b_2=\frac{1}{3+\sqrt{3}}$} & \pbox{20cm}{$c_1=\frac{4}{9}$, $b_1=\frac{1}{2}$\\
        $c_2=5/18$, $b_2=\frac{1}{5-\sqrt{15}}$\\
        $c_3=5/18$, $b_3=\frac{1}{5+\sqrt{15}}$}\\
        \pbox{20cm}{Orders [accumulated]} & $\mathcal{O}(\Delta z_{i,i+1}^3)$ [$\mathcal{O}(\Delta z^2)$] &$\mathcal{O}(\Delta z_{i,i+1}^5)$ [$\mathcal{O}(\Delta z^4)$] & $\mathcal{O}(\Delta z_{i,i+1}^7$ [$\mathcal{O}(\Delta z^6)$] \\
        Speeds up by & $\sim14$ times & $\sim8$ times & $\sim6$ times  
    \end{tabular}
    \end{ruledtabular}
\end{table*}

Here we present a fast algorithm\footnote{The speed comparison is based on a single-core machine.} to compute the above SN Ia likelihood, suitable for data sufficiently compact in the redshift space like the Pantheon compilation. The calculation of $\bm{m_{\rm{p}}}$ involves many evaluations of $f_{\rm{M}}(z)$; each needs an integration over redshift from $0$ to $z$ [see \autoref{eq:comoving-distance-in-H0-flat}]. For the Pantheon compilation, there are $1048$ SNe Ia so that there are $1048$ integrals to evaluate. To reduce computational cost, usually the comoving distance is first evaluated for some redshift grids and then interpolation is used to get the comoving distance at each observed redshift. Here we present an even faster method essentially without a loss of accuracy. This is to take the advantage of the fact that all SN Ia are quite evenly and compactly distributed within $0.01\lesssim z\lesssim 2.3$, using a Runge-Kutta-like algorithm as follows. We first sort the data in a redshift-ascending order and use $i$ to denote the $i$th SN Ia. The redshift separation between every adjoining pair of data points is usually small. This allows us to calculate $f_{\rm{M}}(z_{i+1})$ by adding a small increase to $f_{\rm{M}}(z_i)$. For the first SN Ia with the smallest redshift, we calculate $f_{\rm{M}}(z_1)$ using the integral \autoref{eq:comoving-distance-in-H0-flat}. Then beginning with the second SN Ia, we successively calculate $f_{\rm{M}}(z_{i+1})$ as 
\begin{equation}\label{eq:fM-i-Runge-kuta}
\begin{split}
f_{\rm{M}}(z_{i+1}) &= f_{\rm{M}}(z_{i})+\Delta z_{i,i+1}\sum\limits_{j=1}^n k_j \,, \\
k_j  &= \frac{c_j}{E(z_i+b_j\Delta z_{i,i+1})}\,,
\end{split}
\end{equation}
where $\Delta z_{i,i+1}\equiv z_{i+1}-z_i$ and $E(z)$ is defined in \autoref{eq:Eofz}. In other words, we are replacing each integral for $i\geq2$ by an addition. The number of $k_j$ terms needed, $n$, and the coefficients $c_j$ and $b_j$ are determined by the required order of accuracy. The more terms included in \autoref{eq:fM-i-Runge-kuta}, the more accurate but slower the algorithm becomes. In Table \ref{tab:b-c-coefficients} we list the values of $\bm{c}$ and $\bm{b}$ that allow the error of $f_{\rm{M}}(z_{i+1})-f_{\rm{M}}(z_i)$ to be of the order $\mathcal{O}(\Delta z_{i,i+1}^3)$ for including one term, $\mathcal{O}(\Delta z_{i,i+1}^5)$ for two terms and $\mathcal{O}(\Delta z_{i,i+1}^7)$ for three terms. If $\Delta z_{i,i+1}>0.1$, we force the algorithm to use \autoref{eq:comoving-distance-in-H0-flat} to calculate $f_{\rm{M}}(z_{i+1})$. We tested our algorithm on a python code, and found that it is about an order of magnitude faster than the algorithm that uses \autoref{eq:comoving-distance-in-H0-flat} with the function \textsc{quad} in the \textsc{scipy} package to calculate $f_{\rm{M}}(z)$. While having such a high speed, our algorithm does not lose any accuracy. For the case including two $k$ terms in \autoref{eq:fM-i-Runge-kuta}, the numerical difference ($|\Delta\ln\mathcal{L}|$) between our approximation and the traditional method is well below $10^{-5}$. Considering both accuracy and speed, we use the two-term algorithm in this work. For Pantheon SN Ia compilation, we reproduced $\Omega_{\rm{m}}=0.298\pm0.022$ in the standard $\Lambda$CDM model \citep{Scolnic-etal-2018-Pantheon}.

\bibliography{UCstds}

\providecommand{\noopsort}[1]{}\providecommand{\singleletter}[1]{#1}%
\begin{thebibliography}{}
\expandafter\ifx\csname natexlab\endcsname\relax\def\natexlab#1{#1}\fi
\providecommand{\url}[1]{\href{#1}{#1}}
\providecommand{\dodoi}[1]{doi:~\href{http://doi.org/#1}{\nolinkurl{#1}}}
\providecommand{\doeprint}[1]{\href{http://ascl.net/#1}{\nolinkurl{http://ascl.net/#1}}}
\providecommand{\doarXiv}[1]{\href{https://arxiv.org/abs/#1}{\nolinkurl{https://arxiv.org/abs/#1}}}

\bibitem[{{Abbott} {et~al.}(2017)}]{2017-GW-170717-H0}
{Abbott}, B.~P., {et~al.} 2017, \nat, 551, 85, \dodoi{10.1038/nature24471}

\bibitem[{{Abbott} {et~al.}(2019){Abbott}, others, \& {DES
  Collaboration}}]{2019-DES-year1}
{Abbott}, T.~M.~C., others, \& {DES Collaboration}. 2019, \prd, 99, 123505,
  \dodoi{10.1103/PhysRevD.99.123505}

\bibitem[{Ackermann {et~al.}(2012)}]{Ackermann-etal-science-2012}
Ackermann, M., {et~al.} 2012, Science, 338, 1190,
  \dodoi{10.1126/science.1227160}

\bibitem[{{Addison} {et~al.}(2016){Addison}, {Huang}, {Watts}, {Bennett},
  {Halpern}, {Hinshaw}, \& {Weiland }}]{Addison-etal-2016}
{Addison}, G.~E., {Huang}, Y., {Watts}, D.~J., {et~al.} 2016, \apj, 818, 132,
  \dodoi{10.3847/0004-637X/818/2/132}

\bibitem[{{Agrawal} {et~al.}(2019){Agrawal}, {Cyr-Racine}, {Pinner}, \&
  {Randall}}]{Agrawal-2019lmo}
{Agrawal}, P., {Cyr-Racine}, F.-Y., {Pinner}, D., \& {Randall}, L. 2019, arXiv
  e-prints, arXiv:1904.01016.
\newblock \doarXiv{1904.01016}

\bibitem[{{Aiola} {et~al.}(2020){Aiola}, {Calabrese}, {Maurin}, {Naess},
  {Schmitt}, {Abitbol}, {Addison}, {Ade}, {Alonso}, {Amiri}, {Amodeo},
  {Angile}, {Austermann}, {Baildon}, {Battaglia}, {Beall}, {Bean}, {Becker},
  {Bond}, {Bruno}, {Calafut}, {Campusano}, {Carrero}, {Chesmore}, {Cho},
  {Choi}, {Clark}, {Cothard}, {Crichton}, {Crowley}, {Darwish}, {Datta},
  {Denison}, {Devlin}, {Duell}, {Duff}, {Duivenvoorden}, {Dunkley},
  {D{\"u}nner}, {Essinger-Hileman}, {Fankhanel}, {Ferraro}, {Fox}, {Fuzia},
  {Gallardo}, {Gluscevic}, {Golec}, {Grace}, {Gralla}, {Guan}, {Hall},
  {Halpern}, {Han}, {Hargrave}, {Hasselfield}, {Helton}, {Henderson},
  {Hensley}, {Hill}, {Hilton}, {Hilton}, {Hincks}, {Hlo{\v{z}}ek}, {Ho},
  {Hubmayr}, {Huffenberger}, {Hughes}, {Infante}, {Irwin}, {Jackson}, {Klein},
  {Knowles}, {Koopman}, {Kosowsky}, {Lakey}, {Li}, {Li}, {Li}, {Lokken},
  {Louis}, {Lungu}, {MacInnis}, {Madhavacheril}, {Maldonado}, {Mallaby-Kay},
  {Marsden}, {McMahon}, {Menanteau}, {Moodley}, {Morton}, {Namikawa}, {Nati},
  {Newburgh}, {Nibarger}, {Nicola}, {Niemack}, {Nolta}, {Orlowski-Sherer},
  {Page}, {Pappas}, {Partridge}, {Phakathi}, {Pisano}, {Prince}, {Puddu}, {Qu},
  {Rivera}, {Robertson}, {Rojas}, {Salatino}, {Schaan}, {Schillaci}, {Sehgal},
  {Sherwin}, {Sierra}, {Sievers}, {Sifon}, {Sikhosana}, {Simon}, {Spergel},
  {Staggs}, {Stevens}, {Storer}, {Sunder}, {Switzer}, {Thorne}, {Thornton},
  {Trac}, {Treu}, {Tucker}, {Vale}, {Van Engelen}, {Van Lanen}, {Vavagiakis},
  {Wagoner}, {Wang}, {Ward}, {Wollack}, {Xu}, {Zago}, \&
  {Zhu}}]{2020-Aiola-etal-ACT}
{Aiola}, S., {Calabrese}, E., {Maurin}, L., {et~al.} 2020, \jcap, 2020, 047,
  \dodoi{10.1088/1475-7516/2020/12/047}

\bibitem[{{Alam} {et~al.}(2017)}]{Alam-etal-2017}
{Alam}, S., {et~al.} 2017, \mnras, 470, 2617, \dodoi{10.1093/mnras/stx721}

\bibitem[{{Archidiacono} {et~al.}(2020){Archidiacono}, {Gariazzo}, {Giunti},
  {Hannestad}, \& {Tram}}]{2020-Achidiacono-Gariazzon-Giunti-Hannestad}
{Archidiacono}, M., {Gariazzo}, S., {Giunti}, C., {Hannestad}, S., \& {Tram},
  T. 2020, \jcap, 2020, 029, \dodoi{10.1088/1475-7516/2020/12/029}

\bibitem[{{Asgari} {et~al.}(2021){Asgari}, {Lin}, {Joachimi}, {Giblin},
  {Heymans}, {Hildebrandt}, {Kannawadi}, {St{\"o}lzner}, {Tr{\"o}ster}, {van
  den Busch}, {Wright}, {Bilicki}, {Blake}, {de Jong}, {Dvornik}, {Erben},
  {Getman}, {Hoekstra}, {K{\"o}hlinger}, {Kuijken}, {Miller}, {Radovich},
  {Schneider}, {Shan}, \& {Valentijn}}]{2020-KiDS-1000}
{Asgari}, M., {Lin}, C.-A., {Joachimi}, B., {et~al.} 2021, \aap, 645, A104,
  \dodoi{10.1051/0004-6361/202039070}

\bibitem[{Aylor {et~al.}(2019)Aylor, Joy, Knox, Millea, Raghunathan, \&
  Wu}]{Aylor-etal-2019}
Aylor, K., Joy, M., Knox, L., {et~al.} 2019, The Astrophysical Journal, 874, 4,
  \dodoi{10.3847/1538-4357/ab0898}

\bibitem[{{Baxter} \& {Sherwin}(2020)}]{2020-Baxter-Sherwin}
{Baxter}, E.~J., \& {Sherwin}, B.~D. 2020, \mnras,
  \dodoi{10.1093/mnras/staa3706}

\bibitem[{{Belgacem} {et~al.}(2018){Belgacem}, {Dirian}, {Foffa}, \&
  {Maggiore}}]{Belgacem-etal-2018-nonlocal-MG}
{Belgacem}, E., {Dirian}, Y., {Foffa}, S., \& {Maggiore}, M. 2018, \jcap, 2018,
  002, \dodoi{10.1088/1475-7516/2018/03/002}

\bibitem[{{Bernal} {et~al.}(2021){Bernal}, {Verde}, {Jimenez}, {Kamionkowski},
  {Valcin}, \& {Wandelt}}]{2021-Bernal-etal-new-cosmic-triangle}
{Bernal}, J.~L., {Verde}, L., {Jimenez}, R., {et~al.} 2021, arXiv e-prints,
  arXiv:2102.05066.
\newblock \doarXiv{2102.05066}

\bibitem[{{Beutler} {et~al.}(2011)}]{Beutler-etal-2011}
{Beutler}, F., {et~al.} 2011, \mnras, 416, 3017,
  \dodoi{10.1111/j.1365-2966.2011.19250.x}

\bibitem[{{Birrer} {et~al.}(2019)}]{H0LiCOW-2019}
{Birrer}, S., {et~al.} 2019, \mnras, 484, 4726, \dodoi{10.1093/mnras/stz200}

\bibitem[{{Birrer, S.} {et~al.}(2020){Birrer, S.}, {Shajib, A. J.}, {Galan,
  A.}, {et~al.}}]{Birrer-2020}
{Birrer, S.}, {Shajib, A. J.}, {Galan, A.}, {et~al.} 2020, A\&A, 643, A165,
  \dodoi{10.1051/0004-6361/202038861}

\bibitem[{{Biteau} \& {Williams}(2015)}]{Biteau-Williams-2015}
{Biteau}, J., \& {Williams}, D.~A. 2015, \apj, 812, 60,
  \dodoi{10.1088/0004-637X/812/1/60}

\bibitem[{{Blakeslee} {et~al.}(2021){Blakeslee}, {Jensen}, {Ma}, {Milne}, \&
  {Greene}}]{2021-Blakeslee-etal}
{Blakeslee}, J.~P., {Jensen}, J.~B., {Ma}, C.-P., {Milne}, P.~A., \& {Greene},
  J.~E. 2021, arXiv e-prints, arXiv:2101.02221.
\newblock \doarXiv{2101.02221}

\bibitem[{{Blomqvist} {et~al.}(2019)}]{Blomqvist-etal-2019}
{Blomqvist}, M., {et~al.} 2019, A\&A, 629, A86,
  \dodoi{10.1051/0004-6361/201935641}

\bibitem[{{Blum} {et~al.}(2020){Blum}, {Castorina}, \&
  {Simonovi{\'c}}}]{Blum-etal-2020}
{Blum}, K., {Castorina}, E., \& {Simonovi{\'c}}, M. 2020, \apjl, 892, L27,
  \dodoi{10.3847/2041-8213/ab8012}

\bibitem[{Boddy \& Gluscevic(2018)}]{2018-Boddy-Bluscevic}
Boddy, K.~K., \& Gluscevic, V. 2018, Phys. Rev. D, 98, 083510,
  \dodoi{10.1103/PhysRevD.98.083510}

\bibitem[{{Bond} {et~al.}(1997){Bond}, {Efstathiou}, \&
  {Tegmark}}]{1997-Bond-Efstathiou-Tegmark-1997}
{Bond}, J.~R., {Efstathiou}, G., \& {Tegmark}, M. 1997, \mnras, 291, L33,
  \dodoi{10.1093/mnras/291.1.L33}

\bibitem[{{BOSS Collaboration} {et~al.}(2015){BOSS Collaboration}, {Aubourg},
  {et~al.}}]{2015-Aubourg-BOSS-collaboration}
{BOSS Collaboration}, {Aubourg}, {\'E}., {et~al.} 2015, \prd, 92, 123516,
  \dodoi{10.1103/PhysRevD.92.123516}

\bibitem[{{Boylan-Kolchin} \&
  {Weisz}(2021)}]{2021-Boylan-Kolchin-Weisz-redshift-z-relation}
{Boylan-Kolchin}, M., \& {Weisz}, D.~R. 2021, arXiv e-prints, arXiv:2103.15825.
\newblock \doarXiv{2103.15825}

\bibitem[{{Braglia} {et~al.}(2021){Braglia}, {Ballardini}, {Finelli}, \&
  {Koyama}}]{2020-Braglia-Ballardini-Finelli-Koyama}
{Braglia}, M., {Ballardini}, M., {Finelli}, F., \& {Koyama}, K. 2021, \prd,
  103, 043528, \dodoi{10.1103/PhysRevD.103.043528}

\bibitem[{{Brinckmann} {et~al.}(2020){Brinckmann}, {Hyeok Chang}, \&
  {LoVerde}}]{2020-Brinckmann-Hyeok-Loverde}
{Brinckmann}, T., {Hyeok Chang}, J., \& {LoVerde}, M. 2020, arXiv e-prints,
  arXiv:2012.11830.
\newblock \doarXiv{2012.11830}

\bibitem[{{Chacko} {et~al.}(2016){Chacko}, {Cui}, {Hong}, {Okui}, \&
  {Tsai}}]{2016-Chacho-Cui-Yanou}
{Chacko}, Z., {Cui}, Y., {Hong}, S., {Okui}, T., \& {Tsai}, Y. 2016, Journal of
  High Energy Physics, 2016, 108, \dodoi{10.1007/JHEP12(2016)108}

\bibitem[{{Choi} {et~al.}(2021){Choi}, {Yanagida}, \&
  {Yokozaki}}]{2020-Choi-Yanagida-Yokozaki}
{Choi}, G.~j., {Yanagida}, T.~T., \& {Yokozaki}, N. 2021, Journal of High
  Energy Physics, 2021, 127, \dodoi{10.1007/JHEP01(2021)127}

\bibitem[{{Choudhury} {et~al.}(2021){Choudhury}, {Hannestad}, \&
  {Tram}}]{2020-Choudhury-Hannestad-Tram}
{Choudhury}, S.~R., {Hannestad}, S., \& {Tram}, T. 2021, \jcap, 2021, 084,
  \dodoi{10.1088/1475-7516/2021/03/084}

\bibitem[{{Cooke} {et~al.}(2018){Cooke}, {Pettini}, \&
  {Steidel}}]{Cooke-etal-2018}
{Cooke}, R.~J., {Pettini}, M., \& {Steidel}, C.~C. 2018, \apj, 855, 102,
  \dodoi{10.3847/1538-4357/aaab53}

\bibitem[{{Cuceu} {et~al.}(2019){Cuceu}, {Farr}, {Lemos}, \&
  {Font-Ribera}}]{2019-Cuceu-Farr-Lemos-FontRibera}
{Cuceu}, A., {Farr}, J., {Lemos}, P., \& {Font-Ribera}, A. 2019, \jcap, 2019,
  044, \dodoi{10.1088/1475-7516/2019/10/044}

\bibitem[{{D'Amico} {et~al.}(2021){D'Amico}, {Senatore}, {Zhang}, \&
  {Zheng}}]{DAmico:2020ods}
{D'Amico}, G., {Senatore}, L., {Zhang}, P., \& {Zheng}, H. 2021, \jcap, 2021,
  072, \dodoi{10.1088/1475-7516/2021/05/072}

\bibitem[{{Das} \& {Ghosh}(2020)}]{2020-Das-Ghosh}
{Das}, A., \& {Ghosh}, S. 2020, arXiv e-prints, arXiv:2011.12315.
\newblock \doarXiv{2011.12315}

\bibitem[{{de Jaeger} {et~al.}(2020){de Jaeger}, {Stahl}, {Zheng},
  {Filippenko}, {Riess}, \& {Galbany}}]{2020-Jaeger-etal}
{de Jaeger}, T., {Stahl}, B.~E., {Zheng}, W., {et~al.} 2020, \mnras, 496, 3402,
  \dodoi{10.1093/mnras/staa1801}

\bibitem[{{de Putter} {et~al.}(2019){de Putter}, {Dor{\'e}}, {Gleyzes},
  {Green}, \& {Meyers}}]{2019-dePutter-Dore-Jerome-Green-Meyers}
{de Putter}, R., {Dor{\'e}}, O., {Gleyzes}, J., {Green}, D., \& {Meyers}, J.
  2019, \prl, 122, 041301, \dodoi{10.1103/PhysRevLett.122.041301}

\bibitem[{{DES Collaboration} {et~al.}(2019){DES Collaboration}, {Kessler},
  {et~al.}}]{2019-Kessler-etal}
{DES Collaboration}, {Kessler}, R., {et~al.} 2019, \mnras, 485, 1171,
  \dodoi{10.1093/mnras/stz463}

\bibitem[{{DESI Collaboration} {et~al.}(2016){DESI Collaboration}, {Aghamousa},
  {Aguilar}, {Ahlen}, {Alam}, {Allen}, {Allende Prieto}, {Annis}, {Bailey},
  {Balland}, {Ballester}, {Baltay}, {Beaufore}, {Bebek}, {Beers}, {Bell},
  {Bernal}, {Besuner}, {Beutler}, {Blake}, {Bleuler}, {Blomqvist}, {Blum},
  {Bolton}, {Briceno}, {Brooks}, {Brownstein}, {Buckley-Geer}, {Burden},
  {Burtin}, {Busca}, {Cahn}, {Cai}, {Cardiel-Sas}, {Carlberg}, {Carton},
  {Casas}, {Castander}, {Cervantes-Cota}, {Claybaugh}, {Close}, {Coker},
  {Cole}, {Comparat}, {Cooper}, {Cousinou}, {Crocce}, {Cuby}, {Cunningham},
  {Davis}, {Dawson}, {de la Macorra}, {De Vicente}, {Delubac}, {Derwent},
  {Dey}, {Dhungana}, {Ding}, {Doel}, {Duan}, {Ealet}, {Edelstein},
  {Eftekharzadeh}, {Eisenstein}, {Elliott}, {Escoffier}, {Evatt}, {Fagrelius},
  {Fan}, {Fanning}, {Farahi}, {Farihi}, {Favole}, {Feng}, {Fernandez},
  {Findlay}, {Finkbeiner}, {Fitzpatrick}, {Flaugher}, {Flender}, {Font-Ribera},
  {Forero-Romero}, {Fosalba}, {Frenk}, {Fumagalli}, {Gaensicke}, {Gallo},
  {Garcia-Bellido}, {Gaztanaga}, {Pietro Gentile Fusillo}, {Gerard},
  {Gershkovich}, {Giannantonio}, {Gillet}, {Gonzalez-de-Rivera},
  {Gonzalez-Perez}, {Gott}, {Graur}, {Gutierrez}, {Guy}, {Habib}, {Heetderks},
  {Heetderks}, {Heitmann}, {Hellwing}, {Herrera}, {Ho}, {Holland}, {Honscheid},
  {Huff}, {Hutchinson}, {Huterer}, {Hwang}, {Illa Laguna}, {Ishikawa},
  {Jacobs}, {Jeffrey}, {Jelinsky}, {Jennings}, {Jiang}, {Jimenez}, {Johnson},
  {Joyce}, {Jullo}, {Juneau}, {Kama}, {Karcher}, {Karkar}, {Kehoe}, {Kennamer},
  {Kent}, {Kilbinger}, {Kim}, {Kirkby}, {Kisner}, {Kitanidis}, {Kneib},
  {Koposov}, {Kovacs}, {Koyama}, {Kremin}, {Kron}, {Kronig}, {Kueter-Young},
  {Lacey}, {Lafever}, {Lahav}, {Lambert}, {Lampton}, {Landriau}, {Lang},
  {Lauer}, {Le Goff}, {Le Guillou}, {Le Van Suu}, {Lee}, {Lee}, {Leitner},
  {Lesser}, {Levi}, {L'Huillier}, {Li}, {Liang}, {Lin}, {Linder}, {Loebman},
  {Luki{\'c}}, {Ma}, {MacCrann}, {Magneville}, {Makarem}, {Manera}, {Manser},
  {Marshall}, {Martini}, {Massey}, {Matheson}, {McCauley}, {McDonald},
  {McGreer}, {Meisner}, {Metcalfe}, {Miller}, {Miquel}, {Moustakas}, {Myers},
  {Naik}, {Newman}, {Nichol}, {Nicola}, {Nicolati da Costa}, {Nie}, {Niz},
  {Norberg}, {Nord}, {Norman}, {Nugent}, {O'Brien}, {Oh}, {Olsen}, {Padilla},
  {Padmanabhan}, {Padmanabhan}, {Palanque-Delabrouille}, {Palmese},
  {Pappalardo}, {P{\^a}ris}, {Park}, {Patej}, {Peacock}, {Peiris}, {Peng},
  {Percival}, {Perruchot}, {Pieri}, {Pogge}, {Pollack}, {Poppett}, {Prada},
  {Prakash}, {Probst}, {Rabinowitz}, {Raichoor}, {Ree}, {Refregier}, {Regal},
  {Reid}, {Reil}, {Rezaie}, {Rockosi}, {Roe}, {Ronayette}, {Roodman}, {Ross},
  {Ross}, {Rossi}, {Rozo}, {Ruhlmann-Kleider}, {Rykoff}, {Sabiu}, {Samushia},
  {Sanchez}, {Sanchez}, {Schlegel}, {Schneider}, {Schubnell}, {Secroun},
  {Seljak}, {Seo}, {Serrano}, {Shafieloo}, {Shan}, {Sharples}, {Sholl},
  {Shourt}, {Silber}, {Silva}, {Sirk}, {Slosar}, {Smith}, {Smoot}, {Som},
  {Song}, {Sprayberry}, {Staten}, {Stefanik}, {Tarle}, {Sien Tie}, {Tinker},
  {Tojeiro}, {Valdes}, {Valenzuela}, {Valluri}, {Vargas-Magana}, {Verde},
  {Walker}, {Wang}, {Wang}, {Weaver}, {Weaverdyck}, {Wechsler}, {Weinberg},
  {White}, {Yang}, {Yeche}, {Zhang}, {Zhao}, {Zheng}, {Zhou}, {Zhou}, {Zhu},
  {Zou}, \& {Zu}}]{2016-DESI-white}
{DESI Collaboration}, {Aghamousa}, A., {Aguilar}, J., {et~al.} 2016, arXiv
  e-prints, arXiv:1611.00036.
\newblock \doarXiv{1611.00036}

\bibitem[{{Di Valentino} {et~al.}(2021){Di Valentino}, {Mena}, {Pan},
  {Visinelli}, {Yang}, {Melchiorri}, {Mota}, {Riess}, \&
  {Silk}}]{2021-DiValentino-etal-H0-review}
{Di Valentino}, E., {Mena}, O., {Pan}, S., {et~al.} 2021, arXiv e-prints,
  arXiv:2103.01183.
\newblock \doarXiv{2103.01183}

\bibitem[{{Ding} {et~al.}(2020){Ding}, {Nakama}, \&
  {Wang}}]{2020-Ding-Nakama-Wang-VOID}
{Ding}, Q., {Nakama}, T., \& {Wang}, Y. 2020, Science China Physics, Mechanics,
  and Astronomy, 63, 290403, \dodoi{10.1007/s11433-020-1531-0}

\bibitem[{{Dom{\'\i}nguez} \&
  {Prada}(2013)}]{2013-Dominguez-Prada-gamma-ray-H0}
{Dom{\'\i}nguez}, A., \& {Prada}, F. 2013, \apjl, 771, L34,
  \dodoi{10.1088/2041-8205/771/2/L34}

\bibitem[{Dom{\'{\i}}nguez {et~al.}(2019)Dom{\'{\i}}nguez, Wojtak, Finke,
  Ajello, Helgason, Prada, Desai, Paliya, Marcotulli, \&
  Hartmann}]{Dominguez-etal-2019}
Dom{\'{\i}}nguez, A., Wojtak, R., Finke, J., {et~al.} 2019, The Astrophysical
  Journal, 885, 137, \dodoi{10.3847/1538-4357/ab4a0e}

\bibitem[{Dvorkin {et~al.}(2014)Dvorkin, Blum, \&
  Kamionkowski}]{Dvorkin-Blum-2014}
Dvorkin, C., Blum, K., \& Kamionkowski, M. 2014, Phys. Rev. D, 89, 023519,
  \dodoi{10.1103/PhysRevD.89.023519}

\bibitem[{{eBOSS Collaboration} {et~al.}(2021){eBOSS Collaboration}, {Alam},
  {Aubert}, {Avila}, {Balland}, {Bautista}, {Bershady}, {Bizyaev}, {Blanton},
  {Bolton}, {Bovy}, {Brinkmann}, {Brownstein}, {Burtin}, {Chabanier},
  {Chapman}, {Choi}, {Chuang}, {Comparat}, {Cousinou}, {Cuceu}, {Dawson}, {de
  la Torre}, {de Mattia}, {Agathe}, {des Bourboux}, {Escoffier}, {Etourneau},
  {Farr}, {Font-Ribera}, {Frinchaboy}, {Fromenteau}, {Gil-Mar{\'\i}n}, {Le
  Goff}, {Gonzalez-Morales}, {Gonzalez-Perez}, {Grabowski}, {Guy}, {Hawken},
  {Hou}, {Kong}, {Parker}, {Klaene}, {Kneib}, {Lin}, {Long}, {Lyke}, {de la
  Macorra}, {Martini}, {Masters}, {Mohammad}, {Moon}, {Mueller},
  {Mu{\~n}oz-Guti{\'e}rrez}, {Myers}, {Nadathur}, {Neveux}, {Newman},
  {Noterdaeme}, {Oravetz}, {Oravetz}, {Palanque-Delabrouille}, {Pan}, {Paviot},
  {Percival}, {P{\'e}rez-R{\`a}fols}, {Petitjean}, {Pieri}, {Prakash},
  {Raichoor}, {Ravoux}, {Rezaie}, {Rich}, {Ross}, {Rossi}, {Ruggeri},
  {Ruhlmann-Kleider}, {S{\'a}nchez}, {S{\'a}nchez}, {S{\'a}nchez-Gallego},
  {Sayres}, {Schneider}, {Seo}, {Shafieloo}, {Slosar}, {Smith}, {Stermer},
  {Tamone}, {Tinker}, {Tojeiro}, {Vargas-Maga{\~n}a}, {Variu}, {Wang},
  {Weaver}, {Weijmans}, {Y{\`e}che}, {Zarrouk}, {Zhao}, {Zhao}, \&
  {Zheng}}]{2020-SDSS-IV}
{eBOSS Collaboration}, {Alam}, S., {Aubert}, M., {et~al.} 2021, \prd, 103,
  083533, \dodoi{10.1103/PhysRevD.103.083533}

\bibitem[{{Efstathiou}(2014)}]{2014Efstathiou-Hubble}
{Efstathiou}, G. 2014, \mnras, 440, 1138, \dodoi{10.1093/mnras/stu278}

\bibitem[{{Evslin} {et~al.}(2018){Evslin}, {Sen}, \&
  {Kaushik}}]{Evslin-etal-2018-price-to-shift-H0}
{Evslin}, J., {Sen}, A.~A., \& {Kaushik}, R. 2018, \prd, 97, 103511,
  \dodoi{10.1103/PhysRevD.97.103511}

\bibitem[{{Fields} {et~al.}(2020){Fields}, {Olive}, {Yeh}, \&
  {Young}}]{2020-Fields-Olive-Yeh-Young}
{Fields}, B.~D., {Olive}, K.~A., {Yeh}, T.-H., \& {Young}, C. 2020, \jcap,
  2020, 010, \dodoi{10.1088/1475-7516/2020/03/010}

\bibitem[{{Freedman} {et~al.}(2019){Freedman}, {Madore}, {Hatt}, {Hoyt},
  {Jang}, {Beaton}, {Burns}, {Lee}, {Monson}, {Neeley}, {Phillips}, {Rich}, \&
  {Seibert}}]{Freedman-etal-2019}
{Freedman}, W.~L., {Madore}, B.~F., {Hatt}, D., {et~al.} 2019, \apj, 882, 34,
  \dodoi{10.3847/1538-4357/ab2f73}

\bibitem[{{Freedman} {et~al.}(2020){Freedman}, {Madore}, {Hoyt}, {Jang},
  {Beaton}, {Lee}, {Monson}, {Neeley}, \& {Rich}}]{Freedman-etal-2020}
{Freedman}, W.~L., {Madore}, B.~F., {Hoyt}, T., {et~al.} 2020, \apj, 891, 57,
  \dodoi{10.3847/1538-4357/ab7339}

\bibitem[{{Garcia-Quintero} {et~al.}(2019){Garcia-Quintero}, {Ishak}, {Fox}, \&
  {Lin}}]{2019-Garcia-Ishak-Fox-Lin}
{Garcia-Quintero}, C., {Ishak}, M., {Fox}, L., \& {Lin}, W. 2019, \prd, 100,
  123538, \dodoi{10.1103/PhysRevD.100.123538}

\bibitem[{{Ghosh} {et~al.}(2020){Ghosh}, {Khatri}, \&
  {Roy}}]{2020-Ghosh-Khatri-Roy}
{Ghosh}, S., {Khatri}, R., \& {Roy}, T.~S. 2020, \prd, 102, 123544,
  \dodoi{10.1103/PhysRevD.102.123544}

\bibitem[{{Gomer} \& {Williams}(2020)}]{2019Gomer-etal}
{Gomer}, M., \& {Williams}, L.~L.~R. 2020, \jcap, 2020, 045,
  \dodoi{10.1088/1475-7516/2020/11/045}

\bibitem[{{G{\'o}mez-Valent } \& {Sol{\`a}
  Peracaula}(2018)}]{2018-Gomez-Valent-Sola}
{G{\'o}mez-Valent }, A., \& {Sol{\`a} Peracaula}, J. 2018, \mnras, 478, 126,
  \dodoi{10.1093/mnras/sty1028}

\bibitem[{{Gonzalez} {et~al.}(2020){Gonzalez}, {Hertzberg}, \&
  {Rompineve}}]{2020-Gonzalez-Hertzberg-Rompineve}
{Gonzalez}, M., {Hertzberg}, M.~P., \& {Rompineve}, F. 2020, \jcap, 2020, 028,
  \dodoi{10.1088/1475-7516/2020/10/028}

\bibitem[{Gould \& Schr\'eder(1966)}]{Gould-Schreder-1996}
Gould, R.~J., \& Schr\'eder, G. 1966, Phys. Rev. Lett., 16, 252,
  \dodoi{10.1103/PhysRevLett.16.252}

\bibitem[{{Haridasu} {et~al.}(2021){Haridasu}, {Viel}, \&
  {Vittorio}}]{2020-Haridasu-Balakrishna-Viel}
{Haridasu}, B.~S., {Viel}, M., \& {Vittorio}, N. 2021, \prd, 103, 063539,
  \dodoi{10.1103/PhysRevD.103.063539}

\bibitem[{{H.E.S.S. Collaboration} \& others(2013){H.E.S.S. Collaboration}
  {et~al.}}]{H.E.S.S-collaboration-2013}
{H.E.S.S. Collaboration}, {et~al.} 2013, A\&A, 550, A4,
  \dodoi{10.1051/0004-6361/201220355}

\bibitem[{{Hill} {et~al.}(2020){Hill}, {McDonough}, {Toomey}, \&
  {Alexander}}]{Hill-etal-2020}
{Hill}, J.~C., {McDonough}, E., {Toomey}, M.~W., \& {Alexander}, S. 2020, \prd,
  102, 043507, \dodoi{10.1103/PhysRevD.102.043507}

\bibitem[{{Huang} {et~al.}(2020){Huang}, {Riess}, {Yuan}, {Macri}, {Zakamska},
  {Casertano}, {Whitelock}, {Hoffmann}, {Filippenko}, \&
  {Scolnic}}]{2019-Huang-etal-Mira-H0}
{Huang}, C.~D., {Riess}, A.~G., {Yuan}, W., {et~al.} 2020, \apj, 889, 5,
  \dodoi{10.3847/1538-4357/ab5dbd}

\bibitem[{{Huang} \& {Wang}(2016)}]{2016-Hubble-reconcile}
{Huang}, Q.-G., \& {Wang}, K. 2016, European Physical Journal C, 76, 506,
  \dodoi{10.1140/epjc/s10052-016-4352-x}

\bibitem[{Ivanov {et~al.}(2020)Ivanov, McDonough, Hill,
  Simonovi\ifmmode~\acute{c}\else \'{c}\fi{}, Toomey, Alexander, \&
  Zaldarriaga}]{2020-Ivanov-McDonough-Hill-early-DE-lss}
Ivanov, M.~M., McDonough, E., Hill, J.~C., {et~al.} 2020, Phys. Rev. D, 102,
  103502, \dodoi{10.1103/PhysRevD.102.103502}

\bibitem[{{Jedamzik} \& {Pogosian}(2020)}]{Pogosian-Jedamzik2020}
{Jedamzik}, K., \& {Pogosian}, L. 2020, \prl, 125, 181302,
  \dodoi{10.1103/PhysRevLett.125.181302}

\bibitem[{{Jedamzik} {et~al.}(2021){Jedamzik}, {Pogosian}, \&
  {Zhao}}]{2020-Jedamzik-Pogosian-Zhao}
{Jedamzik}, K., {Pogosian}, L., \& {Zhao}, G.-B. 2021, Communications Physics,
  4, 123, \dodoi{10.1038/s42005-021-00628-x}

\bibitem[{{Jimenez} {et~al.}(2019){Jimenez}, {Cimatti}, {Verde}, {Moresco}, \&
  {Wandelt}}]{2019-Jimenez-etal-stellar-and-cosmic-age}
{Jimenez}, R., {Cimatti}, A., {Verde}, L., {Moresco}, M., \& {Wandelt}, B.
  2019, \jcap, 2019, 043, \dodoi{10.1088/1475-7516/2019/03/043}

\bibitem[{Jimenez \& Loeb(2002)}]{Jimenez-2002-chronometers}
Jimenez, R., \& Loeb, A. 2002, The Astrophysical Journal, 573, 37,
  \dodoi{10.1086/340549}

\bibitem[{{Joudaki} {et~al.}(2020)}]{KiDS-VIKING-DES-2019}
{Joudaki}, S., {et~al.} 2020, \aap, 638, L1,
  \dodoi{10.1051/0004-6361/201936154}

\bibitem[{{Khetan} {et~al.}(2021){Khetan}, {Izzo}, {Branchesi}, {Wojtak},
  {Cantiello}, {Murugeshan}, {Agnello}, {Cappellaro}, {Della Valle}, {Gall},
  {Hjorth}, {Benetti}, {Brocato}, {Burke}, {Hiramatsu}, {Howell}, {Tomasella},
  \& {Valenti}}]{Khetan-etal-2020}
{Khetan}, N., {Izzo}, L., {Branchesi}, M., {et~al.} 2021, \aap, 647, A72,
  \dodoi{10.1051/0004-6361/202039196}

\bibitem[{{Knox} \& {Millea}(2020)}]{2020-Knox-Millea}
{Knox}, L., \& {Millea}, M. 2020, \prd, 101, 043533,
  \dodoi{10.1103/PhysRevD.101.043533}

\bibitem[{Kourkchi {et~al.}(2020)Kourkchi, Tully, Anand, Courtois, Dupuy,
  Neill, Rizzi, \& Seibert}]{2020-Kourkchi-etal}
Kourkchi, E., Tully, R.~B., Anand, G.~S., {et~al.} 2020, The Astrophysical
  Journal, 896, 3, \dodoi{10.3847/1538-4357/ab901c}

\bibitem[{{Kreisch} {et~al.}(2020){Kreisch}, {Cyr-Racine}, \&
  {Dor{\'e}}}]{Kreisch-etal-2019-self-interacting-neutrinos}
{Kreisch}, C.~D., {Cyr-Racine}, F.-Y., \& {Dor{\'e}}, O. 2020, \prd, 101,
  123505, \dodoi{10.1103/PhysRevD.101.123505}

\bibitem[{{Krishnan} {et~al.}(2021){Krishnan}, {{\'O} Colg{\'a}in},
  {Sheikh-Jabbari}, \& {Yang}}]{2020-Krishnan-Colgain-Sheikh}
{Krishnan}, C., {{\'O} Colg{\'a}in}, E., {Sheikh-Jabbari}, M.~M., \& {Yang}, T.
  2021, \prd, 103, 103509, \dodoi{10.1103/PhysRevD.103.103509}

\bibitem[{{Lesgourgues} \& {Pastor}(2012)}]{2012-Lesgourgues-Pastor-nu-mass}
{Lesgourgues}, J., \& {Pastor}, S. 2012, Adv. High Energy Phys., 608515.
\newblock \doarXiv{1212.6154}

\bibitem[{{Lin } \& {Ishak}(2017)}]{WL2017b}
{Lin }, W., \& {Ishak}, M. 2017, Phys. Rev. D, 96, 083532,
  \dodoi{10.1103/PhysRevD.96.083532}

\bibitem[{Lin {et~al.}(2020)Lin, Mack, \& Hou}]{Lin_Mack_Hou_2020}
Lin, W., Mack, K.~J., \& Hou, L. 2020, The Astrophysical Journal, 904, L22,
  \dodoi{10.3847/2041-8213/abc894}

\bibitem[{Lin {et~al.}(2021)}]{2021-Lin-DM}
Lin, W., {et~al.} 2021, unpublished.

\bibitem[{{LSST Dark Energy Science
  Collaboration}(2012)}]{2012-LSST-white-paper}
{LSST Dark Energy Science Collaboration}. 2012, arXiv e-prints,
  arXiv:1211.0310.
\newblock \doarXiv{1211.0310}

\bibitem[{{Maltoni} {et~al.}(2004){Maltoni}, {Schwetz}, {T{\'o}rtola}, \&
  {Valle}}]{2004-Maltoni-Schwetz-nu-masses}
{Maltoni}, M., {Schwetz}, T., {T{\'o}rtola}, M., \& {Valle}, J.~W.~F. 2004, New
  Journal of Physics, 6, 122, \dodoi{10.1088/1367-2630/6/1/122}

\bibitem[{{Millon} {et~al.}(2020){Millon}, {Galan}, {Courbin}, {Treu}, {Suyu},
  {Ding}, {Birrer}, {Chen}, {Shajib}, {Sluse}, {Wong}, {Agnello}, {Auger},
  {Buckley-Geer}, {Chan}, {Collett}, {Fassnacht}, {Hilbert}, {Koopmans},
  {Motta}, {Mukherjee}, {Rusu}, {Sonnenfeld}, {Spiniello}, \& {Van de
  Vyvere}}]{2020-Millon-etal}
{Millon}, M., {Galan}, A., {Courbin}, F., {et~al.} 2020, \aap, 639, A101,
  \dodoi{10.1051/0004-6361/201937351}

\bibitem[{{Moresco} {et~al.}(2020){Moresco}, {Jimenez}, {Verde}, {Cimatti}, \&
  {Pozzetti}}]{2020-Moresco-Raul-Verde}
{Moresco}, M., {Jimenez}, R., {Verde}, L., {Cimatti}, A., \& {Pozzetti}, L.
  2020, \apj, 898, 82, \dodoi{10.3847/1538-4357/ab9eb0}

\bibitem[{{Moresco} \& {Marulli}(2017)}]{2017-Moresco-Marulli-CC}
{Moresco}, M., \& {Marulli}, F. 2017, \mnras, 471, L82,
  \dodoi{10.1093/mnrasl/slx112}

\bibitem[{{Moresco} {et~al.}(2016){Moresco}, {Pozzetti}, {Cimatti}, {Jimenez},
  {Maraston}, {Verde}, {Thomas}, {Citro}, {Tojeiro}, \&
  {Wilkinson}}]{2016-Moresco-etal}
{Moresco}, M., {Pozzetti}, L., {Cimatti}, A., {et~al.} 2016, \jcap, 2016, 014,
  \dodoi{10.1088/1475-7516/2016/05/014}

\bibitem[{{Motloch}(2020)}]{2020a-Motloch}
{Motloch}, P. 2020, \prd, 101, 123509, \dodoi{10.1103/PhysRevD.101.123509}

\bibitem[{Mu\~noz(2019)}]{2019-Munoz-VAO}
Mu\~noz, J.~B. 2019, Phys. Rev. Lett., 123, 131301,
  \dodoi{10.1103/PhysRevLett.123.131301}

\bibitem[{Niedermann \& Sloth(2020)}]{2020-Niedermann-Sloth}
Niedermann, F., \& Sloth, M.~S. 2020, Phys. Rev. D, 102, 063527,
  \dodoi{10.1103/PhysRevD.102.063527}

\bibitem[{O'Malley {et~al.}(2017)O'Malley, Gilligan, \&
  Chaboyer}]{O_Malley-etal-2017}
O'Malley, E.~M., Gilligan, C., \& Chaboyer, B. 2017, The Astrophysical Journal,
  838, 162, \dodoi{10.3847/1538-4357/aa6574}

\bibitem[{{Pan} {et~al.}(2019){Pan}, {Yang}, {Di Valentino}, {Saridakis}, \&
  {Chakraborty}}]{Pan-etal-2019-interacting-DS}
{Pan}, S., {Yang}, W., {Di Valentino}, E., {Saridakis}, E.~N., \&
  {Chakraborty}, S. 2019, \prd, 100, 103520,
  \dodoi{10.1103/PhysRevD.100.103520}

\bibitem[{Park \& Ratra(2019)}]{2018-Park-Ratra}
Park, C.-G., \& Ratra, B. 2019, The Astrophysical Journal, 882, 158,
  \dodoi{10.3847/1538-4357/ab3641}

\bibitem[{{Pesce} {et~al.}(2020){Pesce}, {Braatz}, {Reid}, {Riess}, {Scolnic},
  {Condon}, {Gao}, {Henkel}, {Impellizzeri}, {Kuo}, \& {Lo}}]{Pesce-etal-2020}
{Pesce}, D.~W., {Braatz}, J.~A., {Reid}, M.~J., {et~al.} 2020, \apjl, 891, L1,
  \dodoi{10.3847/2041-8213/ab75f0}

\bibitem[{{Philcox} {et~al.}(2020){Philcox}, {Ivanov}, {Simonovi{\'c}}, \&
  {Zaldarriaga}}]{Philcox-etal-2020}
{Philcox}, O. H.~E., {Ivanov}, M.~M., {Simonovi{\'c}}, M., \& {Zaldarriaga}, M.
  2020, \jcap, 2020, 032, \dodoi{10.1088/1475-7516/2020/05/032}

\bibitem[{{Philcox} {et~al.}(2021){Philcox}, {Sherwin}, {Farren}, \&
  {Baxter}}]{2020-Philcox-Sherwin-Farren-Baxter}
{Philcox}, O. H.~E., {Sherwin}, B.~D., {Farren}, G.~S., \& {Baxter}, E.~J.
  2021, \prd, 103, 023538, \dodoi{10.1103/PhysRevD.103.023538}

\bibitem[{{Planck Collaboration} {et~al.}(2016){Planck Collaboration},
  {Aghanim}, {et~al.}}]{Planck-XI-2016}
{Planck Collaboration}, {Aghanim}, N., {et~al.} 2016, \aap, 594, A11,
  \dodoi{10.1051/0004-6361/201526926}

\bibitem[{{Planck Collaboration} {et~al.}(2020){Planck Collaboration},
  {Aghanim}, {et~al.}}]{2018-Planck-cosmo-params}
---. 2020, \aap, 641, A6, \dodoi{10.1051/0004-6361/201833910}

\bibitem[{{Pogosian} {et~al.}(2020){Pogosian}, {Zhao}, \&
  {Jedamzik}}]{2020-Pogosian-Zhao-Jedamzik}
{Pogosian}, L., {Zhao}, G.-B., \& {Jedamzik}, K. 2020, \apjl, 904, L17,
  \dodoi{10.3847/2041-8213/abc6a8}

\bibitem[{{Poulin} {et~al.}(2019){Poulin}, {Smith}, {Karwal}, \&
  {Kamionkowski}}]{Poulin-etal-2019-EDE}
{Poulin}, V., {Smith}, T.~L., {Karwal}, T., \& {Kamionkowski}, M. 2019, \prl,
  122, 221301, \dodoi{10.1103/PhysRevLett.122.221301}

\bibitem[{{Raveri} {et~al.}(2017){Raveri}, {Hu}, {Hoffman}, \&
  {Wang}}]{2017-Raveri-Hu-Hoffman-Wang-partialy-acoustic-DM}
{Raveri}, M., {Hu}, W., {Hoffman}, T., \& {Wang}, L.-T. 2017, \prd, 96, 103501,
  \dodoi{10.1103/PhysRevD.96.103501}

\bibitem[{{Refsdal}(1964)}]{1964-Refsdal}
{Refsdal}, S. 1964, \mnras, 128, 307, \dodoi{10.1093/mnras/128.4.307}

\bibitem[{{Riess} {et~al.}(2021){Riess}, {Casertano}, {Yuan}, {Bowers},
  {Macri}, {Zinn}, \& {Scolnic}}]{2021-Riess-etal-H0}
{Riess}, A.~G., {Casertano}, S., {Yuan}, W., {et~al.} 2021, \apjl, 908, L6,
  \dodoi{10.3847/2041-8213/abdbaf}

\bibitem[{{Ross} {et~al.}(2015){Ross}, {Samushia}, {Howlett}, {Percival},
  {Burden}, \& {Manera}}]{Ross-etal-2015}
{Ross}, A.~J., {Samushia}, L., {Howlett}, C., {et~al.} 2015, \mnras, 449, 835,
  \dodoi{10.1093/mnras/stv154}

\bibitem[{Sakstein \& Trodden(2020)}]{2020-Jeremy-Mark-early-dark-neutrino}
Sakstein, J., \& Trodden, M. 2020, Phys. Rev. Lett., 124, 161301,
  \dodoi{10.1103/PhysRevLett.124.161301}

\bibitem[{{Schombert} {et~al.}(2020){Schombert}, {McGaugh}, \&
  {Lelli}}]{2020-Schombert-etal}
{Schombert}, J., {McGaugh}, S., \& {Lelli}, F. 2020, \aj, 160, 71,
  \dodoi{10.3847/1538-3881/ab9d88}

\bibitem[{{Scolnic} {et~al.}(2018{\natexlab{a}})}]{Scolnic-etal-2018-Pantheon}
{Scolnic}, D.~M., {et~al.} 2018{\natexlab{a}}, \apj, 859, 101,
  \dodoi{10.3847/1538-4357/aab9bb}

\bibitem[{{Scolnic} {et~al.}(2018{\natexlab{b}})}]{Pantheon-2018}
---. 2018{\natexlab{b}}, \apj, 859, 101, \dodoi{10.3847/1538-4357/aab9bb}

\bibitem[{{Shanks} {et~al.}(2019{\natexlab{a}}){Shanks}, {Hogarth}, \&
  {Metcalfe}}]{2019-Shanks-etal-local-voids1}
{Shanks}, T., {Hogarth}, L.~M., \& {Metcalfe}, N. 2019{\natexlab{a}}, \mnras,
  484, L64, \dodoi{10.1093/mnrasl/sly239}

\bibitem[{{Shanks} {et~al.}(2019{\natexlab{b}}){Shanks}, {Hogarth}, {Metcalfe},
  \& {Whitbourn}}]{2019-Shanks-etal-local-voids2}
{Shanks}, T., {Hogarth}, L.~M., {Metcalfe}, N., \& {Whitbourn}, J.
  2019{\natexlab{b}}, \mnras, 490, 4715, \dodoi{10.1093/mnras/stz2863}

\bibitem[{{Shimon}(2020)}]{2020-Shimon-Weyl-invariant-gravity}
{Shimon}, M. 2020, arXiv e-prints, arXiv:2012.10879.
\newblock \doarXiv{2012.10879}

\bibitem[{Smith {et~al.}(2020)Smith, Poulin, \&
  Amin}]{2020-Smith-Poulin-Amin-oscillating-scalar-field}
Smith, T.~L., Poulin, V., \& Amin, M.~A. 2020, Phys. Rev. D, 101, 063523,
  \dodoi{10.1103/PhysRevD.101.063523}

\bibitem[{{Smith} {et~al.}(2020){Smith}, {Poulin}, {Bernal}, {Boddy},
  {Kamionkowski}, \& {Murgia}}]{2020-Smith-etal}
{Smith}, T.~L., {Poulin}, V., {Bernal}, J.~L., {et~al.} 2020, arXiv e-prints,
  arXiv:2009.10740.
\newblock \doarXiv{2009.10740}

\bibitem[{{Soltis} {et~al.}(2021){Soltis}, {Casertano}, \&
  {Riess}}]{2020-Soltis-Casertano-Riess}
{Soltis}, J., {Casertano}, S., \& {Riess}, A.~G. 2021, \apjl, 908, L5,
  \dodoi{10.3847/2041-8213/abdbad}

\bibitem[{{The LIGO Scientific Collaboration} {et~al.}(2021){The LIGO
  Scientific Collaboration}, {the Virgo Collaboration}, {Abbott}, {Abbott},
  {Abbott}, {Abraham}, {Acernese}, {Ackley}, {Adams}, {Adhikari}, {Adya},
  {Affeldt}, {Agathos}, {Agatsuma}, {Aggarwal}, {Aguiar}, {Aiello}, {Ain},
  {Ajith}, {Allen}, {Allocca}, {Aloy}, {Altin}, {Amato}, {Anand}, {Ananyeva},
  {Anderson}, {Anderson}, {Angelova}, {Antier}, {Appert}, {Arai}, {Araya},
  {Areeda}, {Ar{\`e}ne}, {Arnaud}, {Aronson}, {Arun}, {Ascenzi}, {Ashton},
  {Aston}, {Astone}, {Aubin}, {Aufmuth}, {AultONeal}, {Austin}, {Avendano},
  {Avila-Alvarez}, {Babak}, {Bacon}, {Badaracco}, {Bader}, {Bae}, {Baird},
  {Baker}, {Baldaccini}, {Ballardin}, {Ballmer}, {Bals}, {Banagiri},
  {Barayoga}, {Barbieri}, {Barclay}, {Barish}, {Barker}, {Barkett}, {Barnum},
  {Barone}, {Barr}, {Barsotti}, {Barsuglia}, {Barta}, {Bartlett}, {Bartos},
  {Bassiri}, {Basti}, {Bawaj}, {Bayley}, {Bazzan}, {B{\'e}csy}, {Bejger},
  {Belahcene}, {Bell}, {Beniwal}, {Benjamin}, {Berger}, {Bergmann}, {Bernuzzi},
  {Berry}, {Bersanetti}, {Bertolini}, {Betzwieser}, {Bhandare}, {Bidler},
  {Biggs}, {Bilenko}, {Bilgili}, {Billingsley}, {Birney}, {Birnholtz},
  {Biscans}, {Bischi}, {Biscoveanu}, {Bisht}, {Bitossi}, {Bizouard},
  {Blackburn}, {Blackman}, {Blair}, {Blair}, {Blair}, {Bloemen}, {Bobba},
  {Bode}, {Boer}, {Boetzel}, {Bogaert}, {Bondu}, {Bonnand}, {Booker}, {Boom},
  {Bork}, {Boschi}, {Bose}, {Bossilkov}, {Bosveld}, {Bouffanais}, {Bozzi},
  {Bradaschia}, {Brady}, {Bramley}, {Branchesi}, {Brau}, {Breschi}, {Briant},
  {Briggs}, {Brighenti}, {Brillet}, {Brinkmann}, {Brockill}, {Brooks},
  {Brooks}, {Brown}, {Brunett}, {Buikema}, {Bulik}, {Bulten}, {Buonanno},
  {Buskulic}, {Buy}, {Byer}, {Cabero}, {Cadonati}, {Cagnoli}, {Cahillane},
  {Calder{\'o}n Bustillo}, {Callister}, {Calloni}, {Camp}, {Campbell},
  {Canepa}, {Cannon}, {Cao}, {Cao}, {Carapella}, {Carbognani}, {Caride},
  {Carney}, {Carullo}, {Casanueva Diaz}, {Casentini}, {Caudill},
  {Cavagli{\`a}}, {Cavalier}, {Cavalieri}, {Cella}, {Cerd{\'a}-Dur{\'a}n},
  {Cesarini}, {Chaibi}, {Chakravarti}, {Chamberlin}, {Chan}, {Chao},
  {Charlton}, {Chase}, {Chassande-Mottin}, {Chatterjee}, {Chaturvedi},
  {Cheeseboro}, {Chen}, {Chen}, {Chen}, {Cheng}, {Cheong}, {Chia}, {Chiadini},
  {Chincarini}, {Chiummo}, {Cho}, {Cho}, {Cho}, {Christensen}, {Chu}, {Chua},
  {Chung}, {Chung}, {Ciani}, {Cie{\'s}lar}, {Ciobanu}, {Ciolfi}, {Cipriano},
  {Cirone}, {Clara}, {Clark}, {Clearwater}, {Cleva}, {Coccia}, {Cohadon},
  {Cohen}, {Colleoni}, {Collette}, {Collins}, {Colpi}, {Cominsky},
  {Constancio}, {Conti}, {Cooper}, {Corban}, {Corbitt}, {Cordero-Carri{\'o}n},
  {Corezzi}, {Corley}, {Cornish}, {Corre}, {Corsi}, {Cortese}, {Costa},
  {Cotesta}, {Coughlin}, {Coughlin}, {Coulon}, {Countryman}, {Couvares},
  {Covas}, {Cowan}, {Coward}, {Cowart}, {Coyne}, {Coyne}, {Creighton},
  {Creighton}, {Cripe}, {Croquette}, {Crowder}, {Cullen}, {Cumming},
  {Cunningham}, {Cuoco}, {Dal Canton}, {D{\'a}lya}, {D'Angelo}, {Danilishin},
  {D'Antonio}, {Danzmann}, {Dasgupta}, {Da Silva Costa}, {Datrier}, {Dattilo},
  {Dave}, {Davier}, {Davis}, {Daw}, {DeBra}, {Deenadayalan}, {Degallaix}, {De
  Laurentis}, {Del{\'e}glise}, {Del Pozzo}, {DeMarchi}, {Demos}, {Dent}, {De
  Pietri}, {De Rosa}, {De Rossi}, {DeSalvo}, {de Varona}, {Dhurandhar},
  {D{\'\i}az}, {Dietrich}, {Di Fiore}, {DiFronzo}, {Di Giorgio}, {Di Giovanni},
  {Di Giovanni}, {Di Girolamo}, {Di Lieto}, {Ding}, {Di Pace}, {Di Palma}, {Di
  Renzo}, {Divakarla}, {Dmitriev}, {Doctor}, {Donovan}, {Dooley}, {Doravari},
  {Dorrington}, {Downes}, {Drago}, {Driggers}, {Du}, {Ducoin}, {Dupej},
  {Durante}, {Dwyer}, {Easter}, {Eddolls}, {Edo}, {Effler}, {Ehrens},
  {Eichholz}, {Eikenberry}, {Eisenmann}, {Eisenstein}, {Errico}, {Essick},
  {Estelles}, {Estevez}, {Etienne}, {Etzel}, {Evans}, {Evans}, {Fafone},
  {Fairhurst}, {Fan}, {Farinon}, {Farr}, {Farr}, {Fauchon-Jones}, {Favata},
  {Fays}, {Fazio}, {Fee}, {Feicht}, {Fejer}, {Feng}, {Fernandez-Galiana},
  {Ferrante}, {Ferreira}, {Ferreira}, {Fidecaro}, {Fiori}, {Fiorucci},
  {Fishbach}, {Fisher}, {Fishner}, {Fittipaldi}, {Fitz-Axen}, {Fiumara},
  {Flaminio}, {Fletcher}, {Floden}, {Flynn}, {Fong}, {Font}, {Forsyth},
  {Fournier}, {Hernandez Vivanco}, {Frasca}, {Frasconi}, {Frei}, {Freise},
  {Frey}, {Frey}, {Fritschel}, {Frolov}, {Fronz{\`e}}, {Fulda}, {Fyffe},
  {Gabbard}, {Gadre}, {Gaebel}, {Gair}, {Gammaitoni}, {Gaonkar},
  {Garc{\'\i}a-Quir{\'o}s}, {Garufi}, {Gateley}, {Gaudio}, {Gaur}, {Gayathri},
  {Gemme}, {Genin}, {Gennai}, {George}, {George}, {Gergely}, {Ghonge}, {Ghosh},
  {Ghosh}, {Ghosh}, {Giacomazzo}, {Giaime}, {Giardina}, {Gibson}, {Gill},
  {Glover}, {Gniesmer}, {Godwin}, {Goetz}, {Goetz}, {Goncharov},
  {Gonz{\'a}lez}, {Gonzalez Castro}, {Gopakumar}, {Gossan}, {Gosselin},
  {Gouaty}, {Grace}, {Grado}, {Granata}, {Grant}, {Gras}, {Grassia}, {Gray},
  {Gray}, {Greco}, {Green}, {Green}, {Gretarsson}, {Grimaldi}, {Grimm},
  {Groot}, {Grote}, {Grunewald}, {Gruning}, {Guidi}, {Gulati}, {Guo}, {Gupta},
  {Gupta}, {Gupta}, {Gustafson}, {Gustafson}, {Haegel}, {Halim}, {Hall},
  {Hall}, {Hamilton}, {Hammond}, {Haney}, {Hanke}, {Hanks}, {Hanna}, {Hannam},
  {Hannuksela}, {Hansen}, {Hanson}, {Harder}, {Hardwick}, {Haris}, {Harms},
  {Harry}, {Harry}, {Hasskew}, {Haster}, {Haughian}, {Hayes}, {Healy},
  {Heidmann}, {Heintze}, {Heitmann}, {Hellman}, {Hello}, {Hemming}, {Hendry},
  {Heng}, {Hennig}, {Heurs}, {Hild}, {Hinderer}, {Hochheim}, {Hofman},
  {Holgado}, {Holland}, {Holt}, {Holz}, {Hopkins}, {Horst}, {Hough}, {Howell},
  {Hoy}, {Huang}, {H{\"u}bner}, {Huerta}, {Huet}, {Hughey}, {Hui}, {Husa},
  {Huttner}, {Huynh-Dinh}, {Idzkowski}, {Iess}, {Inchauspe}, {Ingram}, {Inta},
  {Intini}, {Irwin}, {Isa}, {Isac}, {Isi}, {Iyer}, {Jacqmin}, {Jadhav}, {Jani},
  {Janthalur}, {Jaranowski}, {Jariwala}, {Jenkins}, {Jiang}, {Johnson},
  {Jones}, {Jones}, {Jones}, {Jones}, {Jonker}, {Ju}, {Junker}, {Kalaghatgi},
  {Kalogera}, {Kamai}, {Kandhasamy}, {Kang}, {Kanner}, {Kapadia},
  {Karathanasis}, {Karki}, {Kashyap}, {Kasprzack}, {Katsanevas},
  {Katsavounidis}, {Katzman}, {Kaufer}, {Kawabe}, {Keerthana},
  {K{\'e}f{\'e}lian}, {Keitel}, {Kennedy}, {Key}, {Khalili}, {Khan}, {Khan},
  {Khazanov}, {Khetan}, {Khursheed}, {Kijbunchoo}, {Kim}, {Kim}, {Kim}, {Kim},
  {Kim}, {Kim}, {Kimball}, {King}, {Kinley-Hanlon}, {Kirchhoff}, {Kissel},
  {Kleybolte}, {Klika}, {Klimenko}, {Knowles}, {Koch}, {Koehlenbeck},
  {Koekoek}, {Koley}, {Kondrashov}, {Kontos}, {Koper}, {Korobko}, {Korth},
  {Kovalam}, {Kozak}, {Kr{\"a}mer}, {Kringel}, {Krishnendu}, {Kr{\'o}lak},
  {Krupinski}, {Kuehn}, {Kumar}, {Kumar}, {Kumar}, {Kumar}, {Kuo}, {Kutynia},
  {Kwang}, {Lackey}, {Laghi}, {Lai}, {Lam}, {Landry}, {Lane}, {Lang}, {Lange},
  {Lantz}, {Lanza}, {Lartaux-Vollard}, {Lasky}, {Laxen}, {Lazzarini},
  {Lazzaro}, {Leaci}, {Leavey}, {Lecoeuche}, {Lee}, {Lee}, {Lee}, {Lee}, {Lee},
  {Lee}, {Lehmann}, {Lenon}, {Leroy}, {Letendre}, {Levin}, {Li}, {Li}, {Li},
  {Li}, {Li}, {Lin}, {Linde}, {Linker}, {Littenberg}, {Liu}, {Liu},
  {Llorens-Monteagudo}, {Lo}, {London}, {Longo}, {Lorenzini}, {Loriette},
  {Lormand}, {Losurdo}, {Lough}, {Lousto}, {Lovelace}, {Lower}, {L{\"u}ck},
  {Lumaca}, {Lundgren}, {Lynch}, {Ma}, {Macas}, {Macfoy}, {MacInnis},
  {Macleod}, {Macquet}, {Maga{\~n}a Hernandez}, {Maga{\~n}a-Sandoval}, {Magee},
  {Majorana}, {Maksimovic}, {Malik}, {Man}, {Mandic}, {Mangano}, {Mansell},
  {Manske}, {Mantovani}, {Mapelli}, {Marchesoni}, {Marion}, {M{\'a}rka},
  {M{\'a}rka}, {Markakis}, {Markosyan}, {Markowitz}, {Maros}, {Marquina},
  {Marsat}, {Martelli}, {Martin}, {Martin}, {Martinez}, {Martynov},
  {Masalehdan}, {Mason}, {Massera}, {Masserot}, {Massinger}, {Masso-Reid},
  {Mastrogiovanni}, {Matas}, {Matichard}, {Matone}, {Mavalvala}, {McCann},
  {McCarthy}, {McClelland}, {McCormick}, {McCuller}, {McGuire}, {McIsaac},
  {McIver}, {McManus}, {McRae}, {McWilliams}, {Meacher}, {Meadors}, {Mehmet},
  {Mehta}, {Meidam}, {Mejuto Villa}, {Melatos}, {Mendell}, {Mercer}, {Mereni},
  {Merfeld}, {Merilh}, {Merzougui}, {Meshkov}, {Messenger}, {Messick},
  {Messina}, {Metzdorff}, {Meyers}, {Meylahn}, {Miani}, {Miao}, {Michel},
  {Middleton}, {Milano}, {Miller}, {Millhouse}, {Mills}, {Milovich-Goff},
  {Minazzoli}, {Minenkov}, {Mishkin}, {Mishra}, {Mistry}, {Mitra},
  {Mitrofanov}, {Mitselmakher}, {Mittleman}, {Mo}, {Moffa}, {Mogushi},
  {Mohapatra}, {Molina-Ruiz}, {Mondin}, {Montani}, {Moore}, {Moraru},
  {Morawski}, {Moreno}, {Morisaki}, {Mours}, {Mow-Lowry}, {Muciaccia},
  {Mukherjee}, {Mukherjee}, {Mukherjee}, {Mukherjee}, {Mukund}, {Mullavey},
  {Munch}, {Mu{\~n}iz}, {Muratore}, {Murray}, {Nagar}, {Nardecchia},
  {Naticchioni}, {Nayak}, {Neil}, {Neilson}, {Nelemans}, {Nelson}, {Nery},
  {Neunzert}, {Nevin}, {Ng}, {Ng}, {Nguyen}, {Nguyen}, {Nichols}, {Nichols},
  {Nissanke}, {Nocera}, {North}, {Nuttall}, {Obergaulinger}, {Oberling},
  {O'Brien}, {Oganesyan}, {Ogin}, {Oh}, {Oh}, {Ohme}, {Ohta}, {Okada},
  {Oliver}, {Oppermann}, {Oram}, {O'Reilly}, {Ormiston}, {Ortega},
  {O'Shaughnessy}, {Ossokine}, {Ottaway}, {Overmier}, {Owen}, {Pace}, {Pagano},
  {Page}, {Pagliaroli}, {Pai}, {Pai}, {Palamos}, {Palashov}, {Palomba}, {Pan},
  {Panda}, {Pang}, {Pankow}, {Pannarale}, {Pant}, {Paoletti}, {Paoli},
  {Parida}, {Parker}, {Pascucci}, {Pasqualetti}, {Passaquieti}, {Passuello},
  {Patil}, {Patricelli}, {Payne}, {Pearlstone}, {Pechsiri}, {Pedersen},
  {Pedraza}, {Pedurand}, {Pele}, {Penn}, {Perego}, {Perez}, {P{\'e}rigois},
  {Perreca}, {Petermann}, {Pfeiffer}, {Phelps}, {Phukon}, {Piccinni}, {Pichot},
  {Piergiovanni}, {Pierro}, {Pillant}, {Pinard}, {Pinto}, {Pirello}, {Pitkin},
  {Plastino}, {Poggiani}, {Pong}, {Ponrathnam}, {Popolizio}, {Porter},
  {Powell}, {Prajapati}, {Prasad}, {Prasai}, {Prasanna}, {Pratten},
  {Prestegard}, {Principe}, {Prodi}, {Prokhorov}, {Punturo}, {Puppo},
  {P{\"u}rrer}, {Qi}, {Quetschke}, {Quinonez}, {Raab}, {Raaijmakers},
  {Radkins}, {Radulesco}, {Raffai}, {Raja}, {Rajan}, {Rajbhandari},
  {Rakhmanov}, {Ramirez}, {Ramos-Buades}, {Rana}, {Rao}, {Rapagnani},
  {Raymond}, {Razzano}, {Read}, {Regimbau}, {Rei}, {Reid}, {Reitze},
  {Rettegno}, {Ricci}, {Richardson}, {Richardson}, {Ricker}, {Riemenschneider},
  {Riles}, {Rizzo}, {Robertson}, {Robinet}, {Rocchi}, {Rolland}, {Rollins},
  {Roma}, {Romanelli}, {Romano}, {Romano}, {Romel}, {Romie}, {Rose}, {Rose},
  {Rose}, {Rosi{\'n}ska}, {Rosofsky}, {Ross}, {Rowan}, {R{\"u}diger}, {Ruggi},
  {Rutins}, {Ryan}, {Sachdev}, {Sadecki}, {Sakellariadou}, {Salafia},
  {Salconi}, {Saleem}, {Samajdar}, {Sammut}, {Sanchez}, {Sanchez},
  {Sanchis-Gual}, {Sanders}, {Santiago}, {Santos}, {Sarin}, {Sassolas},
  {Sathyaprakash}, {Sauter}, {Savage}, {Schale}, {Scheel}, {Scheuer},
  {Schmidt}, {Schnabel}, {Schofield}, {Sch{\"o}nbeck}, {Schreiber}, {Schulte},
  {Schutz}, {Scott}, {Scott}, {Seidel}, {Sellers}, {Sengupta}, {Sennett},
  {Sentenac}, {Sequino}, {Sergeev}, {Setyawati}, {Shaddock}, {Shaffer},
  {Shahriar}, {Shaner}, {Sharma}, {Sharma}, {Shawhan}, {Shen}, {Shink},
  {Shoemaker}, {Shoemaker}, {Shukla}, {ShyamSundar}, {Siellez}, {Sieniawska},
  {Sigg}, {Singer}, {Singh}, {Singh}, {Singhal}, {Sintes}, {Sitmukhambetov},
  {Skliris}, {Slagmolen}, {Slaven-Blair}, {Smith}, {Smith}, {Somala}, {Son},
  {Soni}, {Sorazu}, {Sorrentino}, {Souradeep}, {Sowell}, {Spencer}, {Spera},
  {Srivastava}, {Srivastava}, {Staats}, {Stachie}, {Standke}, {Steer},
  {Steinke}, {Steinlechner}, {Steinlechner}, {Steinmeyer}, {Stevenson},
  {Stocks}, {Stone}, {Stops}, {Strain}, {Stratta}, {Strigin}, {Strunk},
  {Sturani}, {Stuver}, {Sudhir}, {Summerscales}, {Sun}, {Sunil}, {Sur},
  {Suresh}, {Sutton}, {Swinkels}, {Szczepa{\'n}czyk}, {Tacca}, {Tait},
  {Talbot}, {Tanner}, {Tao}, {T{\'a}pai}, {Tapia}, {Tasson}, {Taylor},
  {Tenorio}, {Terkowski}, {Thomas}, {Thomas}, {Thondapu}, {Thorne}, {Thrane},
  {Tiwari}, {Tiwari}, {Tiwari}, {Toland}, {Tonelli}, {Tornasi},
  {Torres-Forn{\'e}}, {Torrie}, {T{\"o}yr{\"a}}, {Travasso}, {Traylor},
  {Tringali}, {Tripathee}, {Trovato}, {Trozzo}, {Tsang}, {Tse}, {Tso},
  {Tsukada}, {Tsuna}, {Tsutsui}, {Tuyenbayev}, {Ueno}, {Ugolini},
  {Unnikrishnan}, {Urban}, {Usman}, {Vahlbruch}, {Vajente}, {Valdes},
  {Valentini}, {van Bakel}, {van Beuzekom}, {van den Brand}, {Van Den Broeck},
  {Vander-Hyde}, {van der Schaaf}, {VanHeijningen}, {van Veggel}, {Vardaro},
  {Varma}, {Vass}, {Vas{\'u}th}, {Vecchio}, {Vedovato}, {Veitch}, {Veitch},
  {Venkateswara}, {Venugopalan}, {Verkindt}, {Vetrano}, {Vicer{\'e}}, {Viets},
  {Vinciguerra}, {Vine}, {Vinet}, {Vitale}, {Vo}, {Vocca}, {Vorvick},
  {Vyatchanin}, {Wade}, {Wade}, {Wade}, {Walet}, {Walker}, {Wallace}, {Walsh},
  {Wang}, {Wang}, {Wang}, {Wang}, {Wang}, {Ward}, {Warden}, {Warner}, {Was},
  {Watchi}, {Weaver}, {Wei}, {Weinert}, {Weinstein}, {Weiss}, {Wellmann},
  {Wen}, {Wessel}, {We{\ss}els}, {Westhouse}, {Wette}, {Whelan}, {Whiting},
  {Whittle}, {Wilken}, {Williams}, {Williamson}, {Willis}, {Willke}, {Winkler},
  {Wipf}, {Wittel}, {Woan}, {Woehler}, {Wofford}, {Wright}, {Wu}, {Wysocki},
  {Xiao}, {Xu}, {Yamamoto}, {Yancey}, {Yang}, {Yang}, {Yang}, {Yap}, {Yazback},
  {Yeeles}, {Yu}, {Yu}, {Yuen}, {Zadro{\.z}ny}, {Zadro{\.z}ny}, {Zanolin},
  {Zelenova}, {Zendri}, {Zevin}, {Zhang}, {Zhang}, {Zhang}, {Zhao}, {Zhao},
  {Zhou}, {Zhou}, {Zhu}, {Zimmerman}, {Zucker}, {Zweizig}, {LIGO Scientific
  Collaboration}, \& {Virgo Collaboration}}]{2019-GW-BNS-update}
{The LIGO Scientific Collaboration}, {the Virgo Collaboration}, {Abbott},
  B.~P., {et~al.} 2021, \apj, 909, 218, \dodoi{10.3847/1538-4357/abdcb7}

\bibitem[{{Vagnozzi}(2020)}]{Vagnozzi-2019}
{Vagnozzi}, S. 2020, \prd, 102, 023518, \dodoi{10.1103/PhysRevD.102.023518}

\bibitem[{{Valcin} {et~al.}(2021){Valcin}, {Jimenez}, {Verde}, {Bernal}, \&
  {Wandelt}}]{2021-Valcin-etal-cosmic-age}
{Valcin}, D., {Jimenez}, R., {Verde}, L., {Bernal}, J.~L., \& {Wandelt}, B.~D.
  2021, arXiv e-prints, arXiv:2102.04486.
\newblock \doarXiv{2102.04486}

\bibitem[{{Verde} {et~al.}(2019){Verde}, {Treu}, \& {Riess}}]{Verde-etal-2019}
{Verde}, L., {Treu}, T., \& {Riess}, A.~G. 2019, Nature Astronomy, 3, 891,
  \dodoi{10.1038/s41550-019-0902-0}

\bibitem[{{Wagner-Kaiser} {et~al.}(2017){Wagner-Kaiser}, {Sarajedini}, {von
  Hippel}, {Stenning}, {van Dyk}, {Jeffery}, {Robinson}, {Stein}, {Anderson},
  \& {Jefferys}}]{2017-Wagner-Kaiser-et-al}
{Wagner-Kaiser}, R., {Sarajedini}, A., {von Hippel}, T., {et~al.} 2017, \mnras,
  468, 1038, \dodoi{10.1093/mnras/stx544}

\bibitem[{{Wang} {et~al.}(2020)}]{2020-Wang-etc}
{Wang}, Y., {et~al.} 2020, \mnras, \dodoi{10.1093/mnras/staa2593}

\bibitem[{{Xu} {et~al.}(2018){Xu}, {Dvorkin}, \&
  {Chael}}]{Xu-Dvorkin-Chael-2018}
{Xu}, W.~L., {Dvorkin}, C., \& {Chael}, A. 2018, \prd, 97, 103530,
  \dodoi{10.1103/PhysRevD.97.103530}

\bibitem[{{Yang} {et~al.}(2021){Yang}, {Pan}, {Di Valentino}, {Mena}, \&
  {Melchiorri}}]{2021-Yang-Pan-DiValentino-Mena-Melchiorri}
{Yang}, W., {Pan}, S., {Di Valentino}, E., {Mena}, O., \& {Melchiorri}, A.
  2021, arXiv e-prints, arXiv:2101.03129.
\newblock \doarXiv{2101.03129}

\bibitem[{{Yang} {et~al.}(2019){Yang}, {Pan}, {Di Valentino}, {Saridakis}, \&
  {Chakraborty}}]{Yang-eta-2019-many-w}
{Yang}, W., {Pan}, S., {Di Valentino}, E., {Saridakis}, E.~N., \&
  {Chakraborty}, S. 2019, \prd, 99, 043543, \dodoi{10.1103/PhysRevD.99.043543}

\bibitem[{{Yao} {et~al.}(2020){Yao}, {Shan}, {Zhang}, {Kneib}, \&
  {Jullo}}]{2020-Yao-Shan-Zhang-Kneib}
{Yao}, J., {Shan}, H., {Zhang}, P., {Kneib}, J.-P., \& {Jullo}, E. 2020, \apj,
  904, 135, \dodoi{10.3847/1538-4357/abc175}

\bibitem[{{Yu} {et~al.}(2018){Yu}, {Ratra}, \& {Wang}}]{2018Yu-Ratra-Wang}
{Yu}, H., {Ratra}, B., \& {Wang}, F.-Y. 2018, \apj, 856, 3,
  \dodoi{10.3847/1538-4357/aab0a2}

\bibitem[{{Yuan} {et~al.}(2019){Yuan}, {Riess}, {Macri}, {Casertano}, \&
  {Scolnic}}]{Yuan-etal-2019-TRGB-local}
{Yuan}, W., {Riess}, A.~G., {Macri}, L.~M., {Casertano}, S., \& {Scolnic}, D.
  2019, \apj, 886, 61, \dodoi{10.3847/1538-4357/ab4bc9}

\end{thebibliography}
\bibliographystyle{aasjournal}



\end{document}